\documentclass[a4paper,10pt]{revtex4}
\usepackage{mathrsfs}
\usepackage{graphicx}
\usepackage{latexsym}
\usepackage{amsmath}
\usepackage{amssymb}
\usepackage{textcomp}
\usepackage{amsbsy}
\usepackage{graphics}
\usepackage{epstopdf}
\usepackage{color}

\allowdisplaybreaks[4]
\newcommand{\e}{\mathrm{e}}
\newcommand{\nn}{\nonumber \\}

\begin{document}

\title{Unifying Holographic Inflation with Holographic Dark Energy: a Covariant Approach}

\author{Shin'ichi~Nojiri$^{1,2}$\,\thanks{nojiri@gravity.phys.nagoya-u.ac.jp},
S.~D.~Odintsov$^{3,4}$\,\thanks{odintsov@ieec.uab.es},V.~K.~Oikonomou,$^{5,6,7}$\,\thanks{v.k.oikonomou1979@gmail.com},
Tanmoy~Paul$^{8,9}$\,\thanks{pul.tnmy9@gmail.com}} \affiliation{
$^{1)}$ Department of Physics, Nagoya University,
Nagoya 464-8602, Japan \\
$^{2)}$ Kobayashi-Maskawa Institute for the Origin of Particles
and the Universe, Nagoya University, Nagoya 464-8602, Japan \\
$^{3)}$ ICREA, Passeig Luis Companys, 23, 08010 Barcelona, Spain\\
$^{4)}$ Institute of Space Sciences (IEEC-CSIC) C. Can Magrans
s/n, 08193 Barcelona, Spain\\
$^{5)}$ Department of Physics, Aristotle University of
Thessaloniki, Thessaloniki 54124,
Greece\\
$^{6)}$ International Laboratory for Theoretical Cosmology, Tomsk
State University of Control Systems
and Radioelectronics (TUSUR), 634050 Tomsk, Russia\\
$^{7)}$ Tomsk State Pedagogical University, 634061 Tomsk, Russia\\
$^{8)}$ Department of Physics, Chandernagore College, Hooghly - 712 136.\\
$^{9)}$ Department of Theoretical Physics,\\
Indian Association for the Cultivation of Science,\\
2A $\&$ 2B Raja S.C. Mullick Road Kolkata - 700 032, India}

\begin{abstract}
In the present paper, we use the holographic approach  to describe
the early-time acceleration and the late-time acceleration eras of
our Universe in a unified manner. Such ``holographic unification''
is found to have a correspondence with various higher curvature
cosmological models with or without matter fields. The
corresponding holographic cut-offs are determined in terms of the
particle horizon and its derivatives, or the future horizon and
its derivatives. As a result, the holographic energy density we
propose is able to merge various cosmological epochs of the
Universe from a holographic point of view. We find the holographic
correspondence of several $F(R)$ gravity models, including
axion-$F(R)$ gravity models, of several Gauss-Bonnet $F(G)$ models
and finally of $F(T)$ models, and in each case we demonstrate that
it is possible to describe in a unified way inflation and
late-time acceleration in the context of the same holographic
model.
\end{abstract}

%PACS numbers: 04.50.Kd, 95.36.+x, 98.80.-k, 98.80.Cq
%\pacs{04.50.Kd, 95.36.+x, 98.80.-k, 98.80.Cq,11.25.-w}

\maketitle

\section{Introduction}

The holographic principle originates from black hole
thermodynamics and string theory and establishes a connection of
the infrared cutoff of a quantum field theory, which is related to
the vacuum energy, with the largest distance of this theory
\cite{tHooft:1993dmi,Susskind:1994vu,Witten:1998qj,Bousso:2002ju}.
This consideration has been applied extensively in cosmological
considerations, in particular, at the late-time era of the
Universe, known currently as holographic dark energy models
\cite{Li:2004rb,Wang:2016och,Pavon:2005yx,Nojiri:2005pu,Enqvist:2004xv,Zhang:2005yz,Guberina:2005fb,Elizalde:2005ju,
Ito:2004qi,Gong:2004cb,Saridakis:2007cy,Gong:2009dc,BouhmadiLopez:2011xi,Malekjani:2012bw,Khurshudyan:2014axa,Landim:2015hqa,
Gao:2007ep,Li:2008zq,Anagnostopoulos:2020ctz} which are also known to be in good agreement
with observations
\cite{Zhang:2005hs,Li:2009bn,Feng:2007wn,Zhang:2009un,Lu:2009iv,
Micheletti:2009jy,Huang:2004wt,Mukherjee:2017oom}. At this stage
we would like to mention that the most general holographic dark
energy is given by the one with Nojiri-Odintsov cut-off
\cite{Wang:2016och} and it is interesting that it may be applied
to covariant theories, too \cite{Nojiri:2017opc}. Apart from the
dark energy model, the holographic energy density is also found to
be useful to realize the early Universe evolution like the
inflationary evolution
\cite{Horvat:2011wr,Nojiri:2019kkp,Paul:2019hys,Bargach:2019pst,Elizalde:2019jmh,Oliveros:2019rnq}. The first study of 
whether the Higgs inflation respects the holographic principle (within the effective field theoretical approach) was done in \cite{Horvat:2011wr}. 
As a result it was found that the original model of Higgs inflation, where the Higgs field couples with the Ricci scalar, does not 
respect the holographic bound, however a different Higgs inflationary model, where the coupling of the Higgs field is taken with the 
Einstein tensor rather than the Ricci scalar, can change the scenario, in particular this new model passes the holographic test \cite{Horvat:2011wr}. 
In the context of inflation, the holographic model has the 
advantage that since the largest distance (or the cut-off of the
theory) of the early Universe is small, the holographic energy
density is naturally large to successfully trigger the
inflationary era. Moreover the application of the holographic
principle at the early Universe studies, has been extended to the
bouncing scenario by some of our authors in \cite{Nojiri:2019yzg}
where it was shown that the holographic energy density violates
the null energy condition (a necessary condition for bounce
\cite{Lehners:2008vx,Quintin:2014oea,Cai:2013vm,Cai:2016thi,Elizalde:2020zcb}),
which in turn generates the bouncing behavior of the Universe (see
\cite{Brevik:2019mah,Coriano:2019eif} for some more articles on
holographic bounce).

Despite a considerable application of the holographic principle
individually at the early and late time evolution of the Universe,
to date it has not been attempted to provide a unified description
of the inflationary era with the dark energy epoch. In the present
work, we are interested to provide a unified framework of
holographic inflation with holographic dark energy, providing a
unified description of inflation with the late time accelerating
Universe in a holographic context. There are too strong evidence
that eventually modified gravity theories have a prominent role in
describing the early and late-time acceleration eras of our
Universe. Some of the higher curvature models which are well known
to provide such a unified description of early and late-time
acceleration can be found in Refs.
\cite{Nojiri:2010wj,Nojiri:2017ncd,Capozziello:2011et,Artymowski:2014gea,
Nojiri:2003ft,Odintsov:2019mlf,Johnson:2019vwi,Pinto:2018rfg,Odintsov:2019evb,Nojiri:2019riz,Nojiri:2019fft,Lobo:2008sg,
Gorbunov:2010bn,Li:2007xn,Odintsov:2020nwm,Odintsov:2020iui,Appleby:2007vb,Elizalde:2010ts,Cognola:2007zu}
in the context of $F(R)$ gravity,
\cite{Li:2007jm,Odintsov:2018nch,Carter:2005fu,Nojiri:2019dwl,Elizalde:2010jx,Makarenko:2016jsy,delaCruzDombriz:2011wn,Chakraborty:2018scm,
Kanti:2015pda,Kanti:2015dra,Odintsov:2018zhw,Saridakis:2017rdo,Cognola:2006eg}
in the $f(R,\mathcal{G})$ gravity etc. Recently, the axion-$F(R)$
gravity model has been proposed in
\cite{Odintsov:2019evb,Odintsov:2020nwm}, where the axion field
mimics the dark matter evolution and hence the model provides a
description of dark matter along with the unification of early and
late-time acceleration eras. Interestingly, in the present work,
we propose several holographic models which, similar to these
aforementioned higher curvature models, which are able to describe
the inflationary and the dark energy epoch of the Universe in an
unified manner.

The plan of our paper is as follows: in Sec.~\ref{sec_features},
we briefly discuss the essential features of holographic model and
the corresponding holographic cut-off. In the following sections,
we propose various holographic models from different perspective,
which are able to unify the cosmological eras of the Universe.
Finally the conclusions follow in the end of the paper.

\section{Essential features of holographic model}\label{sec_features}

According to the holographic principle, the holographic energy
density is proportional to the inverse squared infrared cutoff
$L_\mathrm{IR}$, which could be related with the causality given
by the cosmological horizon,
\begin{equation}
\label{basic}
\rho_\mathrm{hol}=\frac{3c^2}{\kappa^2 L^2_\mathrm{IR}}\, .
\end{equation}
Here $\kappa^2$ is the gravitational constant and $c$ is a free
parameter. We now consider the Friedmann-Robertson-Walker (FRW)
metric with the flat spatial part,
\begin{equation}
\label{FRWmetric}
ds^2=- dt^2+a^2(t) \sum_{i=1,2,3} \left( dx^i \right)^2 \, ,
\end{equation}
where $a(t)$ is the scale factor. Then the Friedmann equation is
given by,
\begin{equation}
\label{FR1}
H^2=\frac{\kappa^2}{3} \rho\, ,
\end{equation}
where $\rho$ is the energy density of the generalized fluid
driving the expansion of the Universe. We now assume that the
energy density $\rho$ is given by $\rho_\mathrm{hol}$ in
(\ref{basic}). Then the Friedmann equation (\ref{FR1}) can be
rewritten as follows,
\begin{equation}
\label{H2}
H=\frac{c}{L_\mathrm{IR}}\, .
\end{equation}
The infrared cutoff $L_\mathrm{IR}$ is usually assumed to be the
particle horizon $L_\mathrm{p}$ or the future event horizon
$L_\mathrm{f}$, which are given as,
\begin{equation}
\label{H3}
L_\mathrm{p}\equiv a \int_0^t\frac{dt}{a}\ ,\quad
L_\mathrm{f}\equiv a \int_t^\infty \frac{dt}{a}\, .
\end{equation}
Inserting these into (\ref{H2}) we obtain,
\begin{equation}
\label{H5}
\frac{d}{dt}\left(\frac{c}{aH}\right)= \frac{m}{a}\, .
\end{equation}
The $m=1$ case corresponds to the particle horizon and $m=-1$ case
to the future event horizon. In the second case, if we choose
$c=1$, we obtain the solution describing the de Sitter space-time,
\begin{equation}
\label{dS1}
a=a_0\e^{H_0 t}\, ,
\end{equation}
with $a_0$, $H_0$ being two integration constants. Moreover for $c
\neq 1$, Eq. (\ref{H5}) (with $m=-1$) has the solution $a(t) =
a_0\big[t(1-c) + b_0c\big]^{\frac{-c}{1-c}}$, thus we get the
de-Sitter solution only for the case $c=1$. However in the
following sections, when we determine the holographic cut-offs in
terms of the particle or the future horizon, we will keep $c$ as a
free parameter.

In \cite{Nojiri:2005pu}, a general form of the cutoff was proposed,
\begin{equation}
\label{GeneralLIR}
L_\mathrm{IR} = L_\mathrm{IR} \left( L_\mathrm{p}, \dot L_\mathrm{p}, \ddot L_\mathrm{p}, \cdots, L_\mathrm{f}, \dot L_\mathrm{f}, \cdots,
H, \dot H, \cdots R, R_{\mu\nu} R^{\mu\nu}, \cdots \right)\, .
\end{equation}
The above cutoff could be chosen to be equivalent to a general covariant gravity model,
\begin{equation}
\label{GeneralAc}
S = \int d^4 \sqrt{-g} F \left( R,R_{\mu\nu} R^{\mu\nu},
R_{\mu\nu\rho\sigma}R^{\mu\nu\rho\sigma}, \Box R, \Box^{-1} R,
\nabla_\mu R \nabla^\mu R, \cdots \right) \, .
\end{equation}
We will use the above expressions frequently in the following sections.

\section{Holographic correspondence of $F(R)$ gravity without/with matter fields}\label{sec_general_F(R)}

In this section we establish the holographic correspondence of
$F(R)$ gravity with arbitrary form of $F(R)$. By the term
``holographic correspondence'', we mean there exists an equivalent
holographic cut-off ($L_\mathrm{IR}$) which, along with the
expression $H = 1/L_\mathrm{IR}$, can reproduce the cosmological
field equations for the corresponding $F(R)$ gravity. We will show
that such correspondence is not just confined to ``vacuum $F(R)$
models'' but also holds true even for {\it non-vacuum} $F(R)$
models, i.e., the $F(R)$ gravity along with matter fields. We
start with the action of a general $F(R)$ gravity (see
\cite{Nojiri:2010wj,Nojiri:2017ncd,Capozziello:2011et} for general
reviews on $F(R)$ gravity) in absence of matter fields,
\begin{equation}
S = \int d^4x\sqrt{-g} \left[\frac{F(R)}{2\kappa^2}\right]
= \frac{1}{2\kappa^2} \int d^4x\sqrt{-g} \left[R + f(R)\right] \, ,
\label{actionF(R)1}
\end{equation}
where $F(R)$ is decomposed as $F(R) = R + f(R)$ in the second line
and $1/\kappa = M_\mathrm{P}$ with $M_\mathrm{P}$ being the four dimensional
reduced Planck mass. The gravitational field equation for the
above action in the FRW space-time is given by,
\begin{equation}
3H^2 = -\frac{f(R)}{2} + 3\left(H^2 + \dot{H}\right)f'(R) - 3H\frac{df'(R)}{dt} \, ,
\label{eom_F(R)1}
\end{equation}
where $H = \frac{\dot{a}}{a}$ is the Hubble parameter of the
Universe and $f'(R) = \frac{df}{dR}$. Eq.~(\ref{eom_F(R)1}) is the
temporal component of Einstein's field equation, however, the
spatial component, containing $\dot{H}$, can be derived from
Eq.~(\ref{eom_F(R)1}) and thus we do not quote it here. Moreover
in the case of $F(R)$ gravity, the off-diagonal component of
Einstein's field equations are trivial and do not give any new
information for the dynamics of the cosmological equations.
Comparing Eqs.~(\ref{H2}) and (\ref{eom_F(R)1}), we can argue that
the $F(R)$ gravity has a holographic correspondence where the
equivalent holographic cut-off is given by the following
expression,
\begin{equation}
\frac{3c^2}{\left(L_\mathrm{IR}\right)^2} = -\frac{f(R)}{2} + 3\left(H^2 + \dot{H}\right)f'(R) - 18H\left(\ddot{H}
+ 4H\dot{H}\right)f''(R)\, ,
\label{hc1}
\end{equation}
where we express $\frac{df'(R)}{dt} = \dot{R}f''(R) = 6\left(\ddot{H} + 4H\dot{H}\right)f''(R)$ (recall
$R = 12H^2 + 6\dot{H}$ in the FRW space-time). As mentioned
earlier, the holographic cut-off $L_\mathrm{IR}$, in general, is a
function of the particle horizon ($L_\mathrm{p}$), the future
horizon ($L_\mathrm{f}$), the scale factor and their derivatives
(see Eq.~(\ref{GeneralLIR})). Keeping this in mind, here in the
context of $F(R)$ gravity, we determine the holographic cut-off in
two different ways: (1) $L_\mathrm{IR}$ in terms $L_\mathrm{p}$
and their derivatives, and (2) $L_\mathrm{IR}$ in terms of
$L_\mathrm{f}$ and their derivatives. In order to determine the
$L_\mathrm{IR}$ in terms of the particle horizon and their
derivatives, we start from the expression $L_\mathrm{p} =
a\int^{t} \frac{dt}{a(t)}$ as mentioned in Eq.~(\ref{H3}). Upon
differentiating both sides of this expression, one gets the Hubble
parameter as $H(L_\mathrm{p} , \dot{L}_p) =
\frac{\dot{L}_p}{L_\mathrm{p}} - \frac{1}{L_\mathrm{p}}$ which
immediately leads to the Ricci scalar,
\begin{equation}
R^{\left(L_\mathrm{p} \right)} = 6\left[\frac{\ddot{L}_p}{L_\mathrm{p}} + \frac{\dot{L}_p^2}{L_\mathrm{p}^2}
 - \frac{3\dot{L}_p}{L_\mathrm{p}^2} + \frac{2}{L_\mathrm{p}^2}\right] \, .
\label{Riici_particle horizon}
\end{equation}
Plugging the above expressions of the Hubble parameter and the
Ricci scalar into Eq.~(\ref{hc1}), one obtains the $L_\mathrm{IR}
= L_\mathrm{IR}\left(L_\mathrm{p} , \dot{L}_p , \ddot{L}_p ,\
\mbox{higher derivatives of}\ L_\mathrm{p}\right)$ by the
following relation,
\begin{equation}
\frac{3c^2}{\left(L_\mathrm{IR}\right)^2} = -\frac{f \left(R^{\left(L_\mathrm{p} \right)}\right)}{2}
+ 3\left(\frac{\ddot{L}_p}{L_\mathrm{p}} - \frac{\dot{L}_p}{L_\mathrm{p}^2} + \frac{1}{L_\mathrm{p}^2}\right)
f'\left(R^{\left(L_\mathrm{p} \right)} \right)
 - 3\left(\frac{\dot{L}_p}{L_\mathrm{p}} - \frac{1}{L_\mathrm{p}}\right)\frac{df'\left( R^{\left(L_\mathrm{p} \right)} \right)}{dt} \, .
\label{hc_ph_general}
\end{equation}
Similarly, to determine the holographic cut-off as function of the
future horizon ($L_\mathrm{f}$) and their derivatives, we use
$L_\mathrm{f} = a\int_{t}^{\infty} \frac{dt}{a(t)}$. The
derivative on both sides of this expression yields the Hubble
parameter and consequently the Ricci scalar as follows,
\begin{equation}
H(L_\mathrm{f} , \dot{L}_f) = \frac{\dot{L}_f}{L_\mathrm{f}} + \frac{1}{L_\mathrm{f}}\label{Hubble_future horizon} \,,
\end{equation}
and
\begin{equation}
R^{\left(L_\mathrm{f} \right)} = 6\left[\frac{\ddot{L}_f}{L_\mathrm{f}} + \frac{\dot{L}_f^2}{L_\mathrm{f}^2}
+ \frac{3\dot{L}_f}{L_\mathrm{f}^2} + \frac{2}{L_\mathrm{f}^2}\right] \, ,
\label{Ricci_future horizon}
\end{equation}
respectively. Eqs.~(\ref{Hubble_future horizon}) and (\ref{Ricci_future horizon}) along with Eq.~(\ref{hc1}) 
immediately lead to $L_\mathrm{IR} = L_\mathrm{IR}\left(L_\mathrm{f} , \dot{L}_f , \ddot{L}_f ,\ \mbox{higher derivatives of}\ L_\mathrm{f} \right)$ as,
\begin{equation}
\frac{3c^2}{\left(L_\mathrm{IR}\right)^2} 
= -\frac{f \left(R^{\left(L_\mathrm{f} \right)}\right)}{2} 
+ 3\left(\frac{\ddot{L}_f}{L_\mathrm{f}} + \frac{\dot{L}_f}{L_\mathrm{f}^2} 
+ \frac{1}{L_\mathrm{f}^2}\right)f' \left(R^{\left(L_\mathrm{f} \right)} \right)
 - 3\left(\frac{\dot{L}_f}{L_\mathrm{f}} + \frac{1}{L_\mathrm{f}}\right)
 \frac{df' \left(R^{\left(L_\mathrm{f} \right)} \right)}{dt} \, .
\label{hc_fh_general}
\end{equation}
Having established the holographic correspondence of vacuum $F(R)$ model, now we consider the $F(R)$ gravity model
in presence of matter fields and the action is:
\begin{equation}
S = \int d^4x\sqrt{-g} \frac{1}{2\kappa^2}\left[R + f(R) + L_\mathrm{mat}\right] \, ,
\label{actionF(R)2}
\end{equation}
where $L_\mathrm{mat}$ represents the matter field Lagrangian. The
gravitational and the matter field equations for the above action
in FRW spacetime are given by,
\begin{align}
& 3H^2 = -\frac{f(R)}{2} + 3\left(H^2 + \dot{H}\right)f'(R) - 3H\frac{df'(R)}{dt} + \kappa^2 \rho_\mathrm{mat}\, , \nonumber\\
& \dot{\rho}_\mathrm{mat}+ 3H\left(\rho_\mathrm{mat} + p_\mathrm{mat}\right) = 0 \, ,
\label{eom_F(R)_matter1}
\end{align}
with $\rho_\mathrm{mat}$ and $p_\mathrm{mat}$ being the energy
density and pressure of the matter field, respectively. They are
defined as the temporal and spatial component of matter
energy-momentum tensor $T_{\mu\nu} =
\frac{2}{\sqrt{-g}}\frac{\delta L_\mathrm{mat}}{\delta
g^{\mu\nu}}$, respectively. Comparing Eqs.~(\ref{H2}) and
(\ref{eom_F(R)_matter1}), we can consider a holographic origin of
non-vacuum $F(R)$ models where the equivalent holographic cut-off
is given by the same expression as shown in Eq.~(\ref{hc1}).
Consequently, the above gravitational equation turns out to be,
\begin{equation}
3H^2 = \frac{3c^2}{\left(L_\mathrm{IR}\right)^2} + \kappa^2 \rho_\mathrm{mat}\, .
\label{eom_F(R)_matter2}
\end{equation}
Thus Eq.~(\ref{hc_fh_general}) gives the equivalent holographic
cut-off for any arbitrary vacuum/non-vacuum $F(R)$ gravity in
terms of the future horizon and their derivatives, while
Eq.~(\ref{hc_ph_general}) does the same, however in terms of the
particle horizon and their derivatives. These clearly indicate
that $F(R)$ gravity has a holographic origin which mimic the
cosmological field equations of the corresponding $F(R)$ gravity.
In the next sections, we will determine the holographic cut-offs
(by using Eqs.~(\ref{hc_ph_general}) and (\ref{hc_fh_general}))
for some explicit forms of $F(R)$ gravity, which are known to give
a unified picture of inflation and late time dark energy epoch.

At this stage it is worth mentioning that unlike in the two
aforementioned approaches (where $L_\mathrm{IR}$ is determined in
terms of particle horizon or the future horizon), the holographic
cut-offs for various $F(R)$ models are determined in an another
way and in particular, these are determined as an integral, as
shown in Refs. \cite{Nojiri:2019kkp,Paul:2019hys}. With such
integral forms, it is shown that the holographic correspondence
for vacuum and non-vacuum $F(R)$ models exists. In spirit of these
previous works, here we will present integral forms of holographic
cut-offs for the considered forms of $F(R)$ in the next sections.

%%%%%%%%%%%%%%%%%%%%%%%%%%%%%%%%%%%%%%%%%%%%%%%%%%%%%%%%%%%%%%%%%%%%%%%%%%%%%%%%%%%%%%%%%%%%%%%%%%%%%%%%%%%%%%%%%%%%%%%%%%%%%%te

\subsection{Holographic cut-off for $F(R)$ inflationary models}\label{sec_inflationary_F(R)}

Before moving to the unified scenario of inflation and late-time
accelerating epochs, we consider the $F(R)$ inflationary models
and $F(R)$ dark energy models separately and will find the
corresponding holographic cut-offs. The inflationary $F(R)$
holographic cut-offs are shown in this subsection, while the same
for dark energy models are treated in the next subsection. Some of
the popular $F(R)$ models which are known to trigger a viable
inflationary era, are $F(R) = R+\alpha R^2$, $F(R) = \e^{bR}$
(i.e., exponential $F(R)$ gravity) etc. Thus these specific forms
of $F(R)$ are considered here, to determine the holographic
cut-offs. Moreover it has been shown earlier that $F(R)$ models
along with the second rank antisymmetric Kalb-Ramond (KR) fields
also give a viable inflationary era
\cite{Elizalde:2018rmz,Elizalde:2018now}. In fact, the cubic
curvature vacuum $F(R)$ gravity, i.e., $F(R) = R+\beta R^3$ model
does not produce a viable inflation, in particular, the
theoretical expectations of spectral index ($n_s$) and tensor to
scalar ratio ($r$) do not match with the Planck 2018 constraints,
however in the presence of the Kalb-Ramond field the cubic gravity
model becomes compatible with the Planck constraints (i.e., $n_s =
0.9649 \pm 0.0042$ and $r < 0.064$) \cite{Elizalde:2018rmz}. Thus
the impact of the KR field on the inflationary evolution is
significant and will be clear from its cosmological evolution. The
demonstration goes as follows: its EoS parameter is unity, and the
conservation equation of the KR field is given by
$\dot{\rho}_\mathrm{KR} + 6H\rho_\mathrm{KR} = 0$, solving which
one obtains $\rho_\mathrm{KR} \propto 1/a^6$, i.e., the energy
density of the KR field decreases with faster rate in comparison
to pressure-less matter and radiation (the negligible footprint of the KR field in the present universe can also be described 
from the higher dimensional point of view \cite{Paul:2018jpq} where the KR field is 
generally considered to be a bulk field and our four dimensional visible universe is 
a brane embedded within the higher dimensional spacetime, further a non-dynamical approach is also presented in \cite{Das:2018jey} to explain 
the imperceptible signatures of KR field in our universe). Thereby, it clearly depicts
that the present Universe may be free from the direct signatures
of the KR field, however the KR field has considerable effects
during early Universe (when the scale factor is small compared to
the present one). These arguments reveal the importance of the
Kalb-Ramond (KR) field in inflationary models and thus, beside the
vacuum $F(R)$ models, here we also determine the holographic
cut-offs for ``$F(R) + \mathrm{KR}$'' models.

\subsubsection{Quadratic curvature gravity without/with KR field}\label{sec_quadratic_F(R)}

Consider the action $S = \frac{1}{2\kappa^2} \int d^4x\sqrt{-g} \left[ R + \alpha R^2 \right]$
(i.e., quadratic gravity without the KR field and
$\alpha$ is a parameter having mass dimension $[-2]$), for which the Friedmann equation takes the following form,
\begin{equation}
H^2 = 6\alpha \left[\dot{H}^2 - 2H\ddot{H} - 6\dot{H}H^2\right] \, .
\label{quadratic_equation1}
\end{equation}
The equivalent holographic cut-offs for $f(R)=\alpha R^2$ in terms
of $L_\mathrm{p}$ and their derivatives can be obtained from
Eq.~(\ref{hc_ph_general}) and are given by,
\begin{equation}
\frac{c^2}{\left(L_\mathrm{IR}\right)^2} = 6\alpha \left[-\frac{2\dddot{L}_p\dot{L}_p}{L_\mathrm{p}^2}
+ \frac{2\dddot{L}_p}{L_\mathrm{p}^2} + \frac{\ddot{L}_p^2}{L_\mathrm{p}^2} - \frac{2\ddot{L}_p\dot{L}_p^2}{L_\mathrm{p}^3}
+ \frac{6\ddot{L}_p\dot{L}_p}{L_\mathrm{p}^3} - \frac{4\ddot{L}_p}{L_\mathrm{p}^3}
+ \frac{3\dot{L}_p^4}{L_\mathrm{p}^4} - \frac{12\dot{L}_p^3}{L_\mathrm{p}^4}
+ \frac{15\dot{L}_p^2}{L_\mathrm{p}^4} - \frac{6\dot{L}_p}{L_\mathrm{p}^4}\right] \, .
\label{quadratic_ph}
\end{equation}
Similarly by plugging $f(R)=\alpha R^2$ into
Eq.~(\ref{hc_fh_general}), we get the holographic cut-off as a
function of $L_\mathrm{f}$ and their derivatives as follows,
\begin{equation}
\frac{c^2}{\left(L_\mathrm{IR}\right)^2} = 6\alpha \left[-\frac{2\dddot{L}_f\dot{L}_f}{L_\mathrm{f}^2}
 - \frac{2\dddot{L}_f}{L_\mathrm{f}^2} + \frac{\ddot{L}_f^2}{L_\mathrm{f}^2} - \frac{2\ddot{L}_f\dot{L}_f^2}{L_\mathrm{f}^3}
 - \frac{6\ddot{L}_f\dot{L}_f}{L_\mathrm{f}^3} - \frac{4\ddot{L}_f}{L_\mathrm{f}^3} + \frac{3\dot{L}_f^4}{L_\mathrm{f}^4}
+ \frac{12\dot{L}_f^3}{L_\mathrm{f}^4} + \frac{15\dot{L}_f^2}{L_\mathrm{f}^4} + \frac{6\dot{L}_f}{L_\mathrm{f}^4}\right] \, .
\label{quadratic_fh}
\end{equation}
Therefore, the $R^2$ inflationary models which are known to be in
good agreement with observations
\cite{Starobinsky:1980te,Capozziello:2002rd}, have an equivalent
holographic model, thanks to the holographic cut-offs obtained in
Eqs.~(\ref{quadratic_ph}) and (\ref{quadratic_fh}). However, in
\cite{Nojiri:2019kkp}, a different kind of holographic model has
been proposed, where the infrared cut-off takes the following
form,
\begin{equation}
\frac{L_\mathrm{IR}}{c} = -\frac{1}{6\alpha\dot{H}^2a^6} \int dt a^6\dot{H} \, ,
\label{quadratic_integral form1}
\end{equation}
with $\alpha$ being the parameter of the model. Plugging back this
expression into the holographic Friedmann equation $H =
\frac{c}{L_\mathrm{IR}}$, we obtain the cosmological equations for
$F(R) = R + \alpha R^2$ gravity (see
Eq.~(\ref{quadratic_equation1})). Thus, apart from the cut-offs
determined earlier in terms of $L_\mathrm{p}$ or $L_\mathrm{f}$,
the holographic energy density with the cut-off given in
Eq.~(\ref{quadratic_integral form1})) is also able to reproduce
the Starobinsky $R^2$ inflation.

The cut-offs in Eqs.~(\ref{quadratic_ph}), (\ref{quadratic_fh})
can provide an equivalent holographic model even for non-vacuum
quadratic gravity model where the Friedmann equation takes the
form: $3H^2 = \frac{3c^2}{\left(L_\mathrm{IR}\right)^2} +
\kappa^2\rho_\mathrm{mat}$ where $\rho_\mathrm{mat}$ is the matter energy density. We 
consider the Kalb-Ramond field as matter field (keeping the inflationary viability in mind) and the action for ``$R+\alpha
R^2+ \mathrm{KR}$'' model is,
\begin{equation}
S = \int d^4x\sqrt{-g}\left[\frac{1}{2\kappa^2}\left(R + \alpha R^2\right) - \frac{1}{12}H_{\mu\nu\lambda}H^{\mu\nu\lambda}\right] \, ,
\nonumber
\end{equation}
where $H_{\mu\nu\lambda}$ is the field strength tensor of the KR
field defined by $H_{\mu\nu\lambda} =
\partial_{[\mu}B_{\nu\lambda]}$ where $B_{\mu\nu}$ denotes the
second rank antisymmetric KR field. The above model generates a
viable inflationary scenario as explicitly shown in
\cite{Elizalde:2018rmz}. However the KR field indeed affects the
Starobinsky inflationary model by the following ways: being the 
EoS parameter of the KR field is unity, the acceleration of the
early Universe gets reduced due to the presence of the KR field in
comparison to the case where the KR field is absent; moreover from
the observational side, the KR field is found to enhance the
tensor-to-scalar ratio with respect to the Starobinsky
inflationary model. The integral form of the generalized infrared
cut-off for ``$R+\alpha R^2 + \mathrm{KR}$'' model
\cite{Paul:2019hys} is obtained as,
\begin{equation}
\frac{L_\mathrm{IR}}{c} = -\frac{1}{6\alpha\dot{H}^2a^6} \int dt a^6\dot{H} \left(1 - \frac{\kappa^2\rho_0}{a^{6}H^2}\right) \, ,
\label{quadratic_integral form2}
\end{equation}
where $\rho_0$ is the energy density of the KR field at the
horizon crossing. Inserting the above expression into $H =
c/L_\mathrm{IR}$ after some simple algebra, we obtain the
following differential equation,
\begin{equation}
 H^2 = 6\alpha \left[\dot{H}^2 - 2H\ddot{H} - 6\dot{H}H^2\right] + \frac{\kappa^2\rho_0}{a^{6}} \, .
 \label{quadratic_field_equation}
\end{equation}
Eq.~(\ref{quadratic_field_equation}) is actually a combination of
two separate equations,
\begin{equation}
\frac{F(R)}{2} = 3\left( H^2 + \dot{H}\right) F'(R) - 18 \left( 4H^2\dot{H} + H\ddot{H}\right)F''(R)
+ \kappa^2\rho_\mathrm{KR}\, ,
\label{1}
\end{equation}
and
\begin{equation}
 \frac{d\rho_\mathrm{KR}}{dt} + 6H\rho_\mathrm{KR} = 0 \, ,
 \label{2}
\end{equation}
respectively with $F(R) = R + \alpha R^2$. It may be observed that
Eq.~(\ref{2}) is the conservation equation for KR field having EoS
parameter unity. Thereby the cut-off in
Eq.~(\ref{quadratic_integral form2}) can reproduce the
cosmological field equations for ``Starobinsky $+$ KR'' model. It
may be mentioned that for $\rho_0 = 0$, the expression in
Eq.~(\ref{quadratic_integral form2}) becomes the same as in
Eq.~(\ref{quadratic_integral form1}), as expected. Therefore, the
key equations which represent the equivalent holographic energy
density for the $R^2$ model without/with the KR field are given by
Eqs.~(\ref{quadratic_ph}), (\ref{quadratic_fh}),
(\ref{quadratic_integral form1}), and (\ref{quadratic_integral
form2}), respectively.

%%%%%%%%%%%%%
%%%%%%%%%%%%%%

\subsubsection{Cubic curvature gravity without/with KR field}\label{sec_cubic_F(R)}

The cubic curvature model without the KR field has the action $S =
\frac{1}{2\kappa^2} \int d^4x\sqrt{-g}\left[R + \beta R^3\right]$
($\beta$ is a constant parameter with mass dimension $[-4]$)
which, in the Friedmann spacetime, leads to the following
gravitational equation,
\begin{equation}
H^2 = 36\alpha \left[2\dot{H}^3 - 6H\dot{H}\ddot{H} - 15H^2\dot{H}^2 + 4H^6 - 36H^4\dot{H} - 12H^3\ddot{H}\right] \, ,
\label{cubic_equation1}
\end{equation}
Consequently, by plugging the explicit form of $f(R) = \beta R^3$
into Eqs.~(\ref{hc_ph_general}) and (\ref{hc_fh_general}), we
obtain $L_\mathrm{IR} = L_\mathrm{IR}\left(L_\mathrm{p},
\dot{L}_p, \ddot{L}_p, \ \mbox{higher derivatives of}\
L_\mathrm{p} \right)$ and $L_\mathrm{IR} =
L_\mathrm{IR}\left(L_\mathrm{f}, \dot{L}_f, \ddot{L}_f, \
\mbox{higher derivatives of}\ L_\mathrm{f} \right)$ as follows:
\begin{align}
 \frac{c^2}{\left(L_\mathrm{IR}\right)^2}=&\frac{36\alpha}{L_\mathrm{p}^6}
 \left[2 - 3\dot{L}_p + \dot{L}_p^2 + L_\mathrm{p}\ddot{L}_p\right]
\left[2 - 27\dot{L}_p + 59\dot{L}_p^2 - 45\dot{L}_p^3 + 11\dot{L}_p^4 - 13L_\mathrm{p}\ddot{L}_p \right. \nonumber \\
& \left. - L_\mathrm{p}\dot{L}_p^2\ddot{L}_p
+ 2L_\mathrm{p}^2\ddot{L}_p^2 + 6L_\mathrm{p}^2\dddot{L}_p
 + 18L_\mathrm{p}\dot{L}_p\ddot{L}_p - 6L_\mathrm{p}^2\dot{L}_p\dddot{L}_p\right] \, ,
\label{cubic_ph}
\end{align}
and
\begin{align}
\frac{c^2}{\left(L_\mathrm{IR}\right)^2}=&\frac{36\alpha}{L_\mathrm{f}^6}
\left[2 + 3\dot{L}_f + \dot{L}_f^2 + L_\mathrm{f}\ddot{L}_f\right]
\left[2 + 27\dot{L}_f + 59\dot{L}_f^2 + 45\dot{L}_f^3 + 11\dot{L}_f^4 - 13L_\mathrm{f}\ddot{L}_f \right. \nonumber \\
& \left. - L_\mathrm{f}\dot{L}_f^2\ddot{L}_f + 2L_\mathrm{f}^2\ddot{L}_f^2
 - 6L_\mathrm{f}^2\dddot{L}_f
 - 18L_\mathrm{f}\dot{L}_f\ddot{L}_f - 6L_\mathrm{f}^2\dot{L}_f\dddot{L}_f\right] \, ,
 \label{cubic_fh}
\end{align}
respectively. Moreover, the integral form of the holographic
cut-off in the context of vacuum cubic model is obtained as,
\begin{equation}
\frac{L_\mathrm{IR}}{c} = -\frac{\int dt a^{15/2}\dot{H}}
{36\left[2\alpha\dot{H}^3a^{15/2} - 4\alpha H\int dt a^{15/2}H\dot{H}\left(H^3 - 9H\dot{H} - 3\ddot{H}\right)\right]} \, .
\label{cubic_integral form1}
\end{equation}
We can easily see that by inserting the above expression into the
holographic Friedmann equation, we will finally yield
Eq.~(\ref{cubic_equation1}). Therefore, the holographic cut-offs
determined in Eqs.~(\ref{cubic_ph}), (\ref{cubic_fh}), and
(\ref{cubic_integral form1}) are equivalent to $F(R) = R+\beta
R^3$ model. Thereby, the cosmology of $R^3$ gravity can be
realized from holographic origin with the specified infrared
cut-offs just mentioned. It is well known that $F(R) = R + \alpha
R^3$ does not give a viable inflationary era, in particular the
theoretical values of $n_s$ and $r$ do not comply with the
observable constraints from Planck 2018. However, as mentioned
earlier that in presence of the second rank antisymmetric
Kalb-Ramond field, the cubic gravity model becomes viable in
respect to Planck 2018 constraints. In particular for $0.03
\lesssim \kappa^2\rho_0\sqrt{\beta} \lesssim 0.3$ (where $\rho_0$
is the energy density of the KR field at horizon crossing), the
spectral index and the tensor to scalar ratio become
simultaneously compatible with Planck 2018 constraints. In
``$R+\beta R^3+ \mathrm{KR}$'' model, the Friedmann and the KR
field equations are given by,
\begin{align}
0=&-H^2 + 36\alpha \left[2\dot{H}^3 - 6H\dot{H}\ddot{H} - 15H^2\dot{H}^2 + 4H^6 - 36H^4\dot{H} - 12H^3\ddot{H}\right]
+ \kappa^2\rho_\mathrm{KR}\, , \nonumber\\
0=&\frac{d\rho_\mathrm{KR}}{dt} + 6H\rho_\mathrm{KR}\, .
\label{cubic_1}
\end{align}
As demonstrated in Sec.~\ref{sec_general_F(R)}, the holographic
cut-offs for ``$R+\beta R^3+ \mathrm{KR}$'' in terms of
$L_\mathrm{p}$ or $L_\mathrm{f}$ (along with their derivatives)
are same as obtained in Eqs.~(\ref{cubic_ph}) and (\ref{cubic_fh}), 
respectively. On the other hand, the integral form of the
holographic cut-off in ``$R+\beta R^3 + \mathrm{KR}$'' model is
changed in comparison to Eq.~(\ref{cubic_integral form1}) and is
determined as,
\begin{equation}
\frac{L_\mathrm{IR}}{c} = -\frac{\int dt a^{15/2}\dot{H} \left(1 - \frac{\kappa^2h_0}{a^{6}H^2}\right)}
{36\left[2\alpha\dot{H}^3a^{15/2} - 4\alpha H\int dt a^{15/2}H\dot{H}\left(H^3-9H\dot{H}-3\ddot{H}\right)\right]} \, .
\label{cubic_integral form2}
\end{equation}
The above expression along with $H = c/L_\mathrm{IR}$ immediately
leads to the cosmological field Eq.~(\ref{cubic_1}). Again one may
note that for $\rho_0 = 0$, the expression in
Eq.~(\ref{cubic_integral form2}) is reduced to
Eq.~(\ref{cubic_integral form1}), as expected. Thus as a whole,
the cosmological imprints of the cubic gravity model without/with
the KR field can be reproduced by the holographic models having
the cut-offs given in Eqs.~(\ref{cubic_ph}), (\ref{cubic_fh}),
(\ref{cubic_integral form1}), and (\ref{cubic_integral form2}), 
respectively.

%%%%%%%%%%%%%%%%%%%%%%%%%%%%%%%%%%%%%%%%%%%%%%%%%%%%%%%%%%%%%%%%%%%%%%%%%%%%%%%%%%%%%%%%%%%%%%%%%%%%%%%%%%%%%%%%%%%%%%%%%%%ch1

\subsubsection{Exponential $F(R)$ gravity without/with KR field}\label{sec_exp_F(R)}

For exponential $F(R) = a \e^{bR}$ (with $a$ and $b$ are constant
parameters, both having mass dimension $[-2]$), the corresponding
holographic cut-off in terms of $L_\mathrm{p}$ is directly
obtained from Eq.~(\ref{hc_ph_general}) as,
\begin{align}
\frac{c^2}{\left(L_\mathrm{IR}\right)^2}=&\left(ab \e^{bP \left(L_\mathrm{p},\dot{L}_p,\ddot{L}_p \right)}
 - 1\right)\left(\frac{1}{L_\mathrm{p}^2} - \frac{\dot{L}_p}{L_\mathrm{p}^2}
+ \frac{\ddot{L}_p}{L_\mathrm{p}}\right) + \frac{1}{6}\left(P \left(L_\mathrm{p},\dot{L}_p,\ddot{L}_p \right)
 - a \e^{bP \left(L_\mathrm{p},\dot{L}_p,\ddot{L}_p \right)}\right)\nonumber\\
& + 6ab^2 \e^{bP \left(L_\mathrm{p},\dot{L}_p,\ddot{L}_p \right)}\left(\frac{\dot{L}_p}{L_\mathrm{p}}
 - \frac{1}{L_\mathrm{p}}\right)
\left(-4\frac{\dot{L}_p}{L_\mathrm{p}^3} + 6\frac{\dot{L}_p^2}{L_\mathrm{p}^3}
 - 2\frac{\dot{L}_p^3}{L_\mathrm{p}^3} - 3\frac{\ddot{L}_p}{L_\mathrm{p}^2}
+ \frac{\dot{L}_p\ddot{L}_p}{L_\mathrm{p}^2} + \frac{\dddot{L}_p}{L_\mathrm{p}}\right) \, .
\label{exp_ph}
\end{align}
Similarly the cut-off in terms of the future horizon can be
determined from Eq.~(\ref{hc_fh_general}) as,
\begin{align}
\frac{c^2}{\left(L_\mathrm{IR}\right)^2}=&\left(ab \e^{bQ \left(L_\mathrm{f},\dot{L}_f,\ddot{L}_f \right)}
 - 1\right)\left(\frac{1}{L_\mathrm{f}^2} + \frac{\dot{L}_f}{L_\mathrm{f}^2}
+ \frac{\ddot{L}_f}{L_\mathrm{f}}\right) + \frac{1}{6}\left(Q \left(L_\mathrm{f},\dot{L}_f,\ddot{L}_f \right)
 - a \e^{bQ \left(L_\mathrm{f},\dot{L}_f,\ddot{L}_f \right)}\right)\nonumber\\
&+ 6ab^2 \e^{bQ \left(L_\mathrm{f},\dot{L}_f,\ddot{L}_f \right)}\left(\frac{\dot{L}_f}{L_\mathrm{f}}
+ \frac{1}{L_\mathrm{f}}\right)
\left(-4\frac{\dot{L}_f}{L_\mathrm{f}^3} - 6\frac{\dot{L}_f^2}{L_\mathrm{f}^3}
 - 2\frac{\dot{L}_f^3}{L_\mathrm{f}^3} + 3\frac{\ddot{L}_f}{L_\mathrm{f}^2}
+ \frac{\dot{L}_f\ddot{L}_f}{L_\mathrm{f}^2} + \frac{\dddot{L}_f}{L_\mathrm{f}}\right) \, ,
\label{exp_fh}
\end{align}
where the functions $P$ and $Q$ are,
\begin{equation}
P \left(L_\mathrm{p},\dot{L}_p,\ddot{L}_p \right)
= 6\left[\frac{2}{L_\mathrm{p}^2} - \frac{3\dot{L}_p}{L_\mathrm{p}^2} + \frac{\dot{L}_p^2}{L_\mathrm{p}^2}
+ \frac{\ddot{L}_p}{L_\mathrm{p}}\right]\, , \quad
Q \left(L_\mathrm{f},\dot{L}_f,\ddot{L}_f \right)
= 6\left[\frac{2}{L_\mathrm{f}^2} + \frac{3\dot{L}_f}{L_\mathrm{f}^2} + \frac{\dot{L}_f^2}{L_\mathrm{f}^2}
+ \frac{\ddot{L}_f}{L_\mathrm{f}}\right] \, ,
\nonumber
\end{equation}
respectively. Apart from these two kinds of holographic cut-offs,
an integral form of $L_\mathrm{IR}$ in the context of exponential
$F(R)$ gravity is obtained as follows,
\begin{equation}
\frac{L_\mathrm{IR}}{c} = \frac{\int dt a^4\left(1 - 6\beta\dot{H}^2/H^2\right)}
{6\beta a^4\dot{H} + H\int dt a^4\left(\frac{1}{6\beta H^2} - \frac{\dot{H}}{H^2}\right)} \, .
\label{exp_integral form1}
\end{equation}
The above expressions of $L_\mathrm{IR}$ along with the
holographic equation $H = c/L_\mathrm{IR}$ lead to the following
differential equation for $H$:
\begin{equation}
H^2 = 6\beta \left[H\ddot{H} + 4H^2\dot{H}\right] + \frac{1}{6\beta}\left[1 - 6\beta\dot{H}\right] \, ,
\label{exponential_equation}
\end{equation}
which can be rewritten in the form,
\begin{equation}
\frac{F(R)}{2} = 3(H^2 + \dot{H}) F'(R) - 18(4H^2\dot{H} + H\ddot{H})F''(R) \, ,
\nonumber
\end{equation}
with $F(R) \propto \e^{\beta R}$. Therefore the holographic equation can
mimic the cosmological equations of exponential $F(R)$ gravity,
thanks to the different kind of holographic cut-offs in
Eqs.~(\ref{exp_ph}), (\ref{exp_fh}), and (\ref{exp_integral
form1}).

The generalized holographic cut-offs shown in Eqs.~(\ref{exp_ph})
and (\ref{exp_fh}) are also valid for exponential $F(R)$ gravity
even in the presence of the Kalb-Ramond field where the Friedmann
equation takes the following form:
\begin{equation}
 3H^2 = \frac{3c^2}{\left(L_\mathrm{IR}\right)^2} + \kappa^2\rho_\mathrm{KR} \, .\nonumber
\end{equation}
However, in the presence of the KR field, the integral form of $L_\mathrm{IR}$ in the context of
$F(R) = \frac{1}{\kappa^2}\e^{bR}$ comes with the following expression,
\begin{equation}
\frac{L_\mathrm{IR}}{c} = \frac{\int dt a^4\left(1 - 6\beta\dot{H}^2/H^2 - \frac{\kappa^2\rho_0}{a^6H^2}\right)}
{6\beta a^4\dot{H} + H\int dt a^4\left(\frac{1}{6\beta H^2} - \frac{\dot{H}}{H^2}\right)}\, .
\label{exp_integral form2}
\end{equation}
Eqs.~(\ref{exp_ph}), (\ref{exp_fh}), (\ref{exp_integral form1}),
and (\ref{exp_integral form2}) represent different types of
holographic cut-offs which, along with $H=c/L_\mathrm{IR}$,
realize the cosmological scenario of the exponential $F(R)$
gravity without/with the KR field. Regarding the observable
viability, unlike to vacuum cubic curvature gravity, the vacuum
exponential $F(R)$ model is known to be in good agreement with
Planck constraints. Moreover the exponential $F(R)$ model in
presence of the KR field also leads to a viable inflationary
scenario, in particular, for the parametric regime $0.005 \lesssim
\kappa^2\rho_0 b \lesssim 0.1$ (where $\rho_0$ is the KR field
energy density at horizon crossing), the inflationary parameters
like the spectral index and tensor-to-scalar ratio are found to
comply with Planck 2018 constraints \cite{Paul:2019hys}.

\subsection{Holographic cut-off for $F(R)$ dark energy models}\label{sec_DE_F(R)}

As an $F(R)$ dark energy model, we consider,
\begin{equation}
\label{fp1}
f(R) = f_0 R^m \, ,
\end{equation}
with constants $f_0$ and $m$ \cite{Nojiri:2003ft}. We consider the
exponent $m$ to be less than unity for which the term $f_0R^m$
dominates over the Einstein-Hilbert and the matter term(s) in the
low curvature regime, as in the case of late-time Universe. As a
consequence, the Hubble rate $H\equiv \dot a/a$ behaves as,
\begin{equation}
\label{fp2} H \sim \frac{ - \frac{\left( m -1 \right) \left( 2m -
1 \right)}{m-2}}{t} \, ,
\end{equation}
with an effective EoS parameter,
\begin{equation}
w_\mathrm{eff} = -\frac{(6m^2 - 7m - 1)}{3(m - 1)(2m - 1)}\, .
\label{eos_F(R)}
\end{equation}
The above expression indicates that for $m < -0.97$, the
theoretical expectations of the EoS parameter satisfies the
observations coming from SNIa results ($-1.57 < w_\mathrm{eff} <
-0.66$). On other hand, for $m < -0.47$, the EoS parameter in
Eq.~(\ref{eos_F(R)}) satisfies the BAO results ($-2.19 <
w_\mathrm{eff} < -0.42$). Thereby for $m < -0.97$, the theoretical
values of $w_\mathrm{eff}$ is consistent with both the SNIa and
BAO results and thus we stick with $m < -0.97$, for which, our
consideration that the term $R^m$ dominates over the
Einstein-Hilbert term in the low curvature regime is also valid.
Thus $f(R) = f_0R^m$ with $m < -0.97$ can act as a dark energy
model in the context of $F(R)$ gravity. Using
Eqs.~(\ref{hc_ph_general}) and (\ref{hc_fh_general}), we determine
the equivalent holographic cut-off for this model as,
\begin{align}
\label{ftp3}
\frac{3c^2}{\left( L_\mathrm{IR}^{F(R)} \right)^2} = f_0 & \left[ -\frac{1}{2}
\left\{ 6 \left( \frac{\ddot L_\mathrm{p}}{L_\mathrm{p}} + \frac{{\dot L_\mathrm{p}}^2}{L_\mathrm{p}^2}
 - \frac{3\dot L_\mathrm{p}}{L_\mathrm{p}^2} + \frac{2}{L_\mathrm{p}^2} \right) \right\}^m
+ 3 m \left( \frac{\ddot L_\mathrm{p}}{L_\mathrm{p}} - \frac{\dot L_\mathrm{p}}{L_\mathrm{p}^2}
+ \frac{1}{L_\mathrm{p}^2} \right) \left\{ 6 \left( \frac{\ddot L_\mathrm{p}}{L_\mathrm{p}} + \frac{{\dot L_\mathrm{p}}^2}{L_\mathrm{p}^2}
 - \frac{3\dot L_\mathrm{p}}{L_\mathrm{p}^2} + \frac{2}{L_\mathrm{p}^2} \right) \right\}^{m-1} \right. \nn
& \left. - 18m \left( \frac{\dot L_\mathrm{p}}{L_\mathrm{p}} - \frac{1}{L_\mathrm{p}} \right)
\left( \frac{\dddot L_\mathrm{p}}{L_\mathrm{p}} + \frac{\dot L_\mathrm{p} \ddot L_\mathrm{p}}{L_\mathrm{p}^2}
 - \frac{2{\dot L_\mathrm{p}}^3}{L_\mathrm{p}^3} - \frac{3\ddot L_\mathrm{p}}{L_\mathrm{p}^2}
+ \frac{6{ \dot L_\mathrm{p} }^2}{L_\mathrm{p}^3} - \frac{4\dot L_\mathrm{p}}{L_\mathrm{p}^3} \right)
\left\{ 6 \left( \frac{\ddot L_\mathrm{p}}{L_\mathrm{p}} + \frac{{\dot L_\mathrm{p}}^2}{L_\mathrm{p}^2}
 - \frac{3\dot L_\mathrm{p}}{L_\mathrm{p}^2} + \frac{2}{L_\mathrm{p}^2} \right) \right\}^{m-1} \right] \nn
=  f_0 & \left[ -\frac{1}{2}
\left\{ 6 \left( \frac{\ddot L_\mathrm{f}}{L_\mathrm{f}} + \frac{{\dot L_\mathrm{f}}^2}{L_\mathrm{f}^2}
+ \frac{3\dot L_\mathrm{f}}{L_\mathrm{f}^2} + \frac{2}{L_\mathrm{f}^2} \right) \right\}^m
+ 3m \left( \frac{\ddot L_\mathrm{f}}{L_\mathrm{f}} + \frac{\dot L_\mathrm{f}}{L_\mathrm{f}^2}
+ \frac{1}{L_\mathrm{f}^2} \right) \left\{ 6 \left( \frac{\ddot L_\mathrm{f}}{L_\mathrm{f}} + \frac{{\dot L_\mathrm{f}}^2}{L_\mathrm{f}^2}
+ \frac{3\dot L_\mathrm{f}}{L_\mathrm{f}^2} + \frac{2}{L_\mathrm{f}^2} \right) \right\}^{m-1} \right. \nn
& \left. - 18m \left( \frac{\dot L_\mathrm{f}}{L_\mathrm{f}} + \frac{1}{L_\mathrm{f}} \right)
\left( \frac{\dddot L_\mathrm{f}}{L_\mathrm{f}} + \frac{\dot L_\mathrm{f} \ddot L_\mathrm{f}}{L_\mathrm{f}^2}
 - \frac{2{\dot L_\mathrm{f}}^3}{L_\mathrm{f}^3} + \frac{3\ddot L_\mathrm{f}}{L_\mathrm{f}^2}
 - \frac{6{ \dot L_\mathrm{f} }^2}{L_\mathrm{f}^3} - \frac{4\dot L_\mathrm{f}}{L_\mathrm{f}^3} \right)
\left\{ 6 \left( \frac{\ddot L_\mathrm{f}}{L_\mathrm{f}} + \frac{{\dot L_\mathrm{f}}^2}{L_\mathrm{f}^2}
+ \frac{3\dot L_\mathrm{f}}{L_\mathrm{f}^2} + \frac{2}{L_\mathrm{f}^2} \right) \right\}^{m-1} \right] \, .
\end{align}
The first expression in the right hand side of Eq.~(\ref{ftp3})
represents the $L_\mathrm{IR}$ in terms of $L_\mathrm{p}$ and its
derivatives while the second expression gives $L_\mathrm{IR} =
L_\mathrm{IR}\left(L_\mathrm{f} , \dot{L}_f , \ddot{L}_f ,\
\mbox{higher derivatives of}\ L_\mathrm{f} \right)$. Therefore the
model (\ref{fp1}) can also be reproduced from the holographic
energy density having the $L_\mathrm{IR}$ determined in
Eq.~(\ref{ftp3}). Clearly such holographic energy density is able
to drive the dark energy epoch of our Universe. As a more realistic dark energy F(R) model, we may consider the exponential F(R) gravity which 
is known to provide a viable dark energy model as described in \cite{Odintsov:2018qug,Odintsov:2017qif}. In particular, we consider,
\begin{eqnarray}
 F_l(R) = R - 2\Lambda\bigg(1 - e^{-\frac{\beta R}{2\Lambda}}\bigg)
 \label{late_FR2}
\end{eqnarray}
 with $\beta$ and $\Lambda$ are two model parameters 
 having the mass dimensions [0] and [+2] respectively. Due to to the Supernovae 
 Ia (Sne-Ia) \cite{Scolnic:2017caz}, Baryon Accoustic Oscillations (BAO) \cite{Wang:2016wjr}, 
 Cosmic Microwave Background (CMB) \cite{Ade:2015xua} 
 and H(z) \cite{Ratsimbazafy:2017vga} datasets, the parameters 
 $\beta$ and $\Lambda$ are well constrained, in particular, the F(R) model (\ref{late_FR2}) is best fitted with 
 Sne-Ia+BAO+H(z)+CMB data for the parametric regimes given by : $\beta = 3.98^{+\infty}_{-2.46}$ and $\Lambda = 1.2\times10^{-84}$ GeV$^2$ 
 \cite{Odintsov:2018qug}. The equivalent holographic cut-offs (in terms of the particle and future horizon) for the above exponential F(R) model 
 is determined as,
 \begin{align}
\frac{3c^2}{\left( L_\mathrm{IR}^{F(R)} \right)^2} = \Lambda &
\left[ 1 - \exp{\bigg[-\frac{3\beta}{\Lambda}\left(\frac{\ddot{L}_p}{L_\mathrm{p}} + \frac{\dot{L}_p^2}{L_\mathrm{p}^2}
 - \frac{3\dot{L}_p}{L_\mathrm{p}^2} + \frac{2}{L_\mathrm{p}^2}\right)\bigg]} \left\{ 1 
 - \frac{3\beta}{\Lambda} \left( \frac{\ddot L_p}{L_p} - \frac{\dot L_p}{L_p^2}
+ \frac{1}{L_p^2} \right) \right. \right. \nn
& \left. \left.  - \frac{9\beta^2}{\Lambda^2} \left( \frac{\dot L_p}{L_p} - \frac{1}{L_p} \right) 
\left( \frac{\dddot L_p}{L_p} + \frac{\dot L_p \ddot L_p}{L_p^2} 
 - \frac{2{\dot L_p}^3}{L_p^3} - \frac{3\ddot L_p}{L_p^2} 
+ \frac{6{ \dot L_p }^2}{L_p^3} - \frac{4\dot L_p}{L_p^3} \right) \right\} \right] \nn
= \Lambda & \left[ 1 - \exp{\bigg[-\frac{3\beta}{\Lambda}\left(\frac{\ddot{L}_f}{L_\mathrm{f}} + \frac{\dot{L}_f^2}{L_\mathrm{f}^2}
 + \frac{3\dot{L}_f}{L_\mathrm{f}^2} + \frac{2}{L_\mathrm{f}^2}\right)\bigg]} \left\{ 1 
 - \frac{3\beta}{\Lambda} \left( \frac{\ddot L_f}{L_f} + \frac{\dot L_f}{L_f^2}
+ \frac{1}{L_f^2} \right) \right. \right. \nn
& \left. \left.  - \frac{9\beta^2}{\Lambda^2} \left( \frac{\dot L_f}{L_f} + \frac{1}{L_f} \right) 
\left( \frac{\dddot L_f}{L_f} + \frac{\dot L_f \ddot L_f}{L_f^2} 
 - \frac{2{\dot L_f}^3}{L_f^3} + \frac{3\ddot L_f}{L_f^2} 
 - \frac{6{ \dot L_f }^2}{L_f^3} - \frac{4\dot L_f}{L_f^3} \right) \right\} \right] \, .
 \label{expf2}
\end{align}
Because the $F(R)$ gravity model (\ref{late_FR2}) generates the accelerating expansion of the 
late universe, the holographic model also plays the role of the dark energy.

\subsection{Unification of holographic inflation with holographic dark energy}\label{sec_unification_F(R)}

In view of the previous sections, we are motivated to construct a
model unifying the inflationary era in terms of the Starobinsky
$R^2$ inflation (which is known to lead to inflationary
observables in a very good agreement with observations
\cite{Akrami:2018odb}), with the accelerating expansion of the
late Universe by using the future event horizon $L_\mathrm{f}$ in
(\ref{H3}). An example is given by,
\begin{equation}
\label{HDEu1}
\frac{L_\mathrm{IR}}{c} = \frac{a^{1+n}}{\left( a_0^\mathrm{l} \right)^n + a^n} \int_t^\infty \frac{dt}{a}
 - \frac{\left(a_0^\mathrm{e}\right)^m}{6\alpha{\dot H}^2 a^6
\left( \left( a_0^\mathrm{e}\right) ^m + a^m \right)}
\int^t dt a^{6} \dot H \, .
\end{equation}
Here $n$, $m$, $a_0^\mathrm{l}$, and $a_0^\mathrm{e}$ are positive
constants and we choose $a_0^\mathrm{l}$ to be smaller than the
scale factor in the present Universe and $a_0^\mathrm{e}$ to be
larger the scale factor in the Universe after the inflation. We
also assume $a_0^\mathrm{l} \gg a_0^\mathrm{e}$. Then in the late
Universe, where $a\gg a_0^\mathrm{l}$, the first term dominates
and behaves as the future horizon,
\begin{equation}
\label{HDEu2}
\frac{L_\mathrm{IR}}{c} \sim a \int_t^\infty \frac{dt}{a} \, ,
\end{equation}
which along with the holographic Friedmann equation $H =
\frac{c}{L_{IR}}$ generates the accelerating expansion of the
present Universe. On the other hand, in the early Universe, where
$a\ll a_0^\mathrm{e}$, the second term dominates and behaves as in
(\ref{quadratic_integral form1}),
\begin{equation}
\label{HDEu3}
\frac{L_\mathrm{IR}}{c}  \sim - \frac{1}{6\alpha{\dot H}^2 a^6}  \int^t dt a^{6} \dot H \, ,
\end{equation}
which generates the Starobinsky inflation. Thereby the cut-off
proposed in Eq.~(\ref{HDEu1}) can provide an unified scenario of
inflation and late time acceleration of the Universe from
holographic point of view.

\subsubsection{Minimally coupled axion-$F(R)$ gravity model}

A more realistic and a recent model which unifies various
cosmological epochs of the Universe is the axion-$F(R)$ gravity
model described in \cite{Odintsov:2020nwm} and first proposed in
\cite{Odintsov:2019evb}. The action of the model is,
\begin{equation}
S = \int d^4x\sqrt{-g}\left[\frac{1}{2\kappa^2}\left(R + f(R)\right) - \frac{1}{2}\partial_{\mu}\phi\partial^{\mu}\phi - V(\phi)
+ L_\mathrm{mat}\right] \, ,
\label{axion1_action}
\end{equation}
where $\phi$ is the axion scalar field endowed within the
potential $V(\phi)$. The axion field acts as a dark matter
component of the Universe, which during the inflationary era, was
frozen at its primordial vacuum expectation value.
$L_\mathrm{mat}$ is the matter Lagrangian, however, in
\cite{Odintsov:2020nwm}, the authors assumed the only perfect
fluid to be present is the radiation fluid, which in fact, comes
as a viable consideration for the purpose of unification. Here it deserves mentioning that the model (\ref{axion1_action}) does not 
describe the interaction between the ordinary/dark matter and the dark energy. However, in the next subsection in the model (\ref{axion2_action}), 
the axion field i.e the dark matter component is considered to be non-minimally coupled with the curvature. The form 
of $f(R)$ of action (\ref{axion1_action}) is taken as,
\begin{equation}
f(R) = \frac{R^2}{M^2} - \gamma R^{\delta} \, ,
\label{axion1_form of FR}
\end{equation}
with $\delta$ is a positive number in the interval $0 < \delta <
1$. Moreover the parameter $M$ is chosen as $M = 1.5\times
10^{-5}\left(\frac{N}{50}\right)^{-1}$ for early-time
phenomenological reasons \cite{Odintsov:2020nwm} where $N$ is the
e-foldings number. The first Friedmann equation of the action
(\ref{axion1_action}) is,
\begin{equation}
3H^2 = -\frac{f(R)}{2} + 3(H^2 + \dot{H})f'(R) - 3H\frac{df'(R)}{dt} + \kappa^2\left(\rho_\mathrm{r} + \frac{1}{2}\dot{\phi}^2
+ V(\phi)\right) \, ,
 \label{axion1_eom}
\end{equation}
with $\rho_\mathrm{r}$ being the energy density for the radiation
(recall $L_\mathrm{mat}$ consists only of radiation, as mentioned
earlier) and the scalar field is considered to be spatially
homogeneous. As described in \cite{Odintsov:2020nwm}, the model
(\ref{axion1_action}) successfully unifies various epochs of our
Universe. The demonstration goes as follows : during the early
epoch when the curvature is large, the term $R^2$ dominates over
$R^{\delta}$ as $0 < \delta < 1$. Also in the early stage of the
Universe, the axion field was frozen in its vacuum expectation
value and for $m_{\phi} \sim \mathcal{O}
\left(10^{-14}\right)\,\mathrm{eV}$, the axion field only
contributes a very small cosmological constant (compared to the
other terms) in the equation of motion. The axion mass in the present model (i.e $\sim \mathcal{O}
\left(10^{-14}\right)\,\mathrm{eV}$) respects the latest Planck data of the dark matter density given by $\Omega_ah^2 = 0.12\pm0.001$. This result 
is in agreement with \cite{Marsh:2015xka} where it was shown that 
the dark matter density (i.e $\Omega_ah^2 = 0.12\pm0.001$) requires the axion mass range as 
$10^{-24} \mathrm{eV} \leq m_{\phi} \leq 10^{-12} \mathrm{eV}$. for $10^{14} \mathrm{GeV} \lesssim \phi_i \lesssim 10^{17} \mathrm{GeV}$, where $\phi_i$ 
being the vacuum expectation value obtained by the axion field. Further 
the ``string anthropic boundary`` where $\Omega_ah^2 = 0.12$ leads to $m_{\phi} \simeq 10^{-19} \mathrm{eV}$ for 
$\phi_i = 10^{16} \mathrm{eV}$ \cite{Marsh:2015xka}. 
Here we would like to mention that in the present paper we discuss the Axion Like Particles ALP (where the Peccei Quinn symmetry is not 
broken during inflation), not the QCD axion in which case the mass bound of axion 
lie within $10^{-5} \mathrm{eV} \lesssim m_{\phi} \lesssim 10^{-2} \mathrm{eV}$ \cite{Ringwald:2016yge}. 
Thus the the mass range relevant for axion dark matter is wide, 
as demonstrated in various earlier literatures. For the present higher curvature axion model, the axion mass comes 
as $m_{\phi} \sim \mathcal{O}\left(10^{-14}\right)\,\mathrm{eV}$ with $\phi_i \sim \mathcal{O}\left(10^{15}\right)\,\mathrm{GeV}$. 
Coming back to the Eq.(\ref{axion1_eom}), it is evident that the Ricci scalar related terms dominate the inflationary
evolution, and specifically the $R^2$, hence the model is reduced
to the $R^2$ model, which yields a viable inflationary
phenomenology compatible with the observational data coming from
Planck 2018. With the expansion of the Universe, the Hubble
parameter decreases and when $H \lesssim m_{\phi}$, the axion
starts to oscillate. Assuming a slowly varying oscillation for the
axion, it can be shown that the axion energy density scales as
$\rho_{\phi} \sim a^{-3}$, thus the axion mimics the dark matter
fluid with an average EoS parameter $w_{\phi} \simeq 0$. At late
time of the Universe, the $R^{\delta}$ term in the $f(R)$
dominates and controls the dynamics. After demonstrating the
contribution of each term, the full Friedmann equation is solved
numerically for a wide range of redshift ($z$), in particular for
$z = [0,10]$. Following the numerical solution, various
parameters, namely the deceleration parameter $q = -1 -
\frac{\dot{H}}{H^2}$, the jerk parameter $j = \frac{\ddot{H}}{H^3}
- 3q - 2$, the parameter $s = \frac{(j-1)}{3(q - 1/2)}$, and the
parameter $Om(z) = \frac{\frac{H(z)}{H_0^2} - 1}{(1+z)^3 - 1}$
have been estimated. As a result the axion-$F(R)$ gravity model is
found to produce results very similar to the $\Lambda$CDM model,
in some cases almost identical for small redshifts, and in all
cases compatible results with the latest Planck constraints on the
cosmological parameters.

In order to map the axion-$F(R)$ gravity model with the
holographic one, we put the form of $f(R) = \frac{R^2}{M^2} -
\gamma R^{\delta}$ into Eq.~(\ref{hc_ph_general}) and upon some
simple algebra, we get the corresponding holographic cut-off in
terms of $L_\mathrm{p}$ and its derivatives as follows:
\begin{align}
\frac{c^2}{\left(L_\mathrm{IR}\right)^2}=&6\left(\frac{1}{L_\mathrm{p}} - \frac{\dot{L}_p}{L_\mathrm{p}}\right)
\left[\frac{2}{M^2} + \left(6\Omega_{\left(L_\mathrm{p} \right)}\right)^{\delta-2}\gamma\delta(1-\delta)\right]
\left[-4\frac{\dot{L}_p}{L_\mathrm{p}^3} + 6\frac{\dot{L}_p^2}{L_\mathrm{p}^3}
 - 2\frac{\dot{L}_p^3}{L_\mathrm{p}^3} - 3\frac{\ddot{L}_p}{L_\mathrm{p}^2} + \frac{\dot{L}_p\ddot{L}_p}{L_\mathrm{p}^2}
+ \frac{\dddot{L}_p}{L_\mathrm{p}}\right]\nonumber\\
&+6\Omega_{\left(L_\mathrm{p} \right)}\left[\frac{1}{L_\mathrm{p}^2} - \frac{\dot{L}_p}{L_\mathrm{p}^2}
+ \frac{\ddot{L}_p}{L_\mathrm{p}}\right]
\left[\frac{2}{M^2} - \left(6\Omega_{\left(L_\mathrm{p} \right)}\right)^{\delta-2}\gamma\delta\right]
 - 6\Omega_{\left(L_\mathrm{p} \right)}\left[\frac{2}{M^2} - \left(6\Omega_{\left(L_\mathrm{p} \right)}\right)^{\delta-2}\gamma\right] \, ,
\label{axion1_hc_ph}
\end{align}
with $\Omega_{\left(L_\mathrm{p} \right)} =
\frac{2}{L_\mathrm{p}^2} - 3\frac{\dot{L}_p}{L_\mathrm{p}^2} +
\frac{\dot{L}_p^2}{L_\mathrm{p}^2} +
\frac{\ddot{L}_p}{L_\mathrm{p}}$. Similarly the cut-off in terms
of the $L_\mathrm{f}$ and its derivatives is obtained from
Eq.~(\ref{hc_fh_general}) as follows,
\begin{align}
\frac{c^2}{\left(L_\mathrm{IR}\right)^2}=&6\left(-\frac{1}{L_\mathrm{f}} - \frac{\dot{L}_f}{L_\mathrm{f}}\right)
\left[\frac{2}{M^2} + \left(6\Omega_{\left(L_\mathrm{f} \right)}\right)^{\delta-2}\gamma\delta(1-\delta)\right]
\left[-4\frac{\dot{L}_f}{L_\mathrm{f}^3} - 6\frac{\dot{L}_f^2}{L_\mathrm{f}^3} - 2\frac{\dot{L}_f^3}{L_\mathrm{f}^3}
+ 3\frac{\ddot{L}_f}{L_\mathrm{f}^2} + \frac{\dot{L}_f\ddot{L}_f}{L_\mathrm{f}^2}
+ \frac{\dddot{L}_f}{L_\mathrm{f}}\right]\nonumber\\
&+6\Omega_{\left(L_\mathrm{f} \right)}\left[\frac{1}{L_\mathrm{f}^2} + \frac{\dot{L}_f}{L_\mathrm{f}^2}
+ \frac{\ddot{L}_f}{L_\mathrm{f}}\right]
\left[\frac{2}{M^2} - \left(6\Omega_{\left(L_\mathrm{f} \right)}\right)^{\delta-2}\gamma\delta\right]
 - 6\Omega_{\left(L_\mathrm{f} \right)}\left[\frac{2}{M^2} - \left(6\Omega_{\left(L_\mathrm{f} \right)}\right)^{\delta-2}\gamma\right] \, ,
\label{axion1_hc_fh}
\end{align}
where $\Omega_{\left(L_\mathrm{f} \right)} = \frac{2}{L_\mathrm{p}^2} + 3\frac{\dot{L}_f}{L_\mathrm{f}^2}
+ \frac{\dot{L}_f^2}{L_\mathrm{f}^2} + \frac{\ddot{L}_f}{L_\mathrm{f}}$.
With the cut-offs determined in the above two expressions, the axion-$F(R)$ model (\ref{axion1_action})
can be equivalently mapped to a holographic model
where the Friedmann equation is of the form:
\begin{equation}
3H^2 = \frac{3c^2}{\left(L_\mathrm{IR}\right)^2} + \kappa^2\left(\rho_\mathrm{r} + \frac{1}{2}\dot{\phi}^2 + V(\phi)\right) \, .
\label{holographic_friedmann}
\end{equation}
The cut-offs determined in Eqs.~(\ref{axion1_hc_ph}) and
(\ref{axion1_hc_fh}) can be decomposed as,
\begin{equation}
\frac{1}{\left(L_\mathrm{IR}\right)^2} = \frac{1}{\left(L^{(1)}_\mathrm{IR}\right)^2}
+ \frac{1}{\left(L^{(2)}_\mathrm{IR}\right)^2}\, , \nonumber
\end{equation}
or the above expression can be rewritten as,
\begin{equation}
\rho_\mathrm{hol} = \rho_\mathrm{hol}^{(1)} + \rho_\mathrm{hol}^{(2)} \, ,
\label{axion1_decomposition}
\end{equation}
where $\rho_\mathrm{hol}^{(i)} \left(= \frac{3c^2}{\kappa^2\left(L^{(i)}_\mathrm{IR}\right)^2}\right)$
is the holographic energy density with the cut-off $L^{(i)}_\mathrm{IR}$.
Furthermore  $\rho_\mathrm{hol}^{(1)}$ and $\rho_\mathrm{hol}^{(2)}$ are given by,
\begin{align}
&\rho_\mathrm{hol}^{(1)}=\frac{3c^2}{\kappa^2\left(L^{(1)}_\mathrm{IR}\right)^2}\nonumber\\
=&\frac{3}{\kappa^2}\bigg\{\frac{12}{M^2}\left(\frac{1}{L_\mathrm{p}} - \frac{\dot{L}_p}{L_\mathrm{p}}\right)
\left[-4\frac{\dot{L}_p}{L_\mathrm{p}^3} + 6\frac{\dot{L}_p^2}{L_\mathrm{p}^3}
 - 2\frac{\dot{L}_p^3}{L_\mathrm{p}^3} - 3\frac{\ddot{L}_p}{L_\mathrm{p}^2} + \frac{\dot{L}_p\ddot{L}_p}{L_\mathrm{p}^2}
+ \frac{\dddot{L}_p}{L_\mathrm{p}}\right] + \frac{12}{M^2}\Omega_{\left(L_\mathrm{p} \right)}\left[\frac{1}{L_\mathrm{p}^2}
 - \frac{\dot{L}_p}{L_\mathrm{p}^2} + \frac{\ddot{L}_p}{L_\mathrm{p}}\right]
 - \frac{12}{M^2}\Omega_{\left(L_\mathrm{p} \right)}\bigg\} \nonumber\\
=&\frac{3}{\kappa^2}\bigg\{\frac{12}{M^2}\left(-\frac{1}{L_\mathrm{f}} - \frac{\dot{L}_f}{L_\mathrm{f}}\right)
 \left[-4\frac{\dot{L}_f}{L_\mathrm{f}^3} - 6\frac{\dot{L}_f^2}{L_\mathrm{f}^3} - 2\frac{\dot{L}_f^3}{L_\mathrm{f}^3} + 3\frac{\ddot{L}_f}{L_\mathrm{f}^2} + \frac{\dot{L}_f\ddot{L}_f}{L_\mathrm{f}^2}
+ \frac{\dddot{L}_f}{L_\mathrm{f}}\right]
+ \frac{12}{M^2}\Omega_{\left(L_\mathrm{f} \right)}\left[\frac{1}{L_\mathrm{f}^2} + \frac{\dot{L}_f}{L_\mathrm{f}^2} + \frac{\ddot{L}_f}{L_\mathrm{f}}\right]  - \frac{12}{M^2}\Omega_{\left(L_\mathrm{f} \right)}\bigg\} \, ,
 \label{axion1_cut-off_early}
\end{align}
and
\begin{align}
\rho_\mathrm{hol}^{(2)} = \frac{3c^2}{\kappa^2\left(L^{(2)}_\mathrm{IR}\right)^2}
=&\frac{3}{\kappa^2}\bigg\{6\left(6\Omega_{\left(L_\mathrm{p} \right)}\right)^{\delta-2}\gamma\delta(1-\delta)\left(\frac{1}{L_\mathrm{p}}
 - \frac{\dot{L}_p}{L_\mathrm{p}}\right)
\left[-4\frac{\dot{L}_p}{L_\mathrm{p}^3} + 6\frac{\dot{L}_p^2}{L_\mathrm{p}^3} - 2\frac{\dot{L}_p^3}{L_\mathrm{p}^3}
 - 3\frac{\ddot{L}_p}{L_\mathrm{p}^2} + \frac{\dot{L}_p\ddot{L}_p}{L_\mathrm{p}^2}
+ \frac{\dddot{L}_p}{L_\mathrm{p}}\right]\nonumber\\
&-\left(6\Omega_{\left(L_\mathrm{p} \right)}\right)^{\delta-1}\gamma\delta\left[\frac{1}{L_\mathrm{p}^2}
 - \frac{\dot{L}_p}{L_\mathrm{p}^2} + \frac{\ddot{L}_p}{L_\mathrm{p}}\right]
 + \gamma\left(6\Omega_{\left(L_\mathrm{p} \right)}\right)^{\delta-1}\bigg\} \nonumber\\
=&\frac{3}{\kappa^2}\bigg\{16\left(6\Omega_{\left(L_\mathrm{f} \right)}\right)^{\delta-2}\gamma\delta(1-\delta)\left(-\frac{1}{L_\mathrm{f}}
 - \frac{\dot{L}_f}{L_\mathrm{f}}\right)
\left[-4\frac{\dot{L}_f}{L_\mathrm{f}^3} - 6\frac{\dot{L}_f^2}{L_\mathrm{f}^3}
 - 2\frac{\dot{L}_f^3}{L_\mathrm{f}^3} + 3\frac{\ddot{L}_f}{L_\mathrm{f}^2}+ \frac{\dot{L}_f\ddot{L}_f}{L_\mathrm{f}^2}
+ \frac{\dddot{L}_f}{L_\mathrm{f}}\right]\nonumber\\
&-\left(6\Omega_{\left(L_\mathrm{f} \right)}\right)^{\delta-1}\gamma\delta\left[\frac{1}{L_\mathrm{f}^2}
+ \frac{\dot{L}_f}{L_\mathrm{f}^2} + \frac {\ddot{L}_f}{L_\mathrm{f}}\right]
+ \gamma\left(6\Omega_{\left(L_\mathrm{f} \right)}\right)^{\delta-1}\bigg\} \, ,
\label{axion1_cut-off_late}
\end{align}
respectively. With such decomposition of $L_\mathrm{IR}$, the
holographic Friedmann Eq.~(\ref{holographic_friedmann}) can be
rewritten as,
\begin{equation}
3H^2 = \kappa^2\left(\rho_\mathrm{hol}^{(1)} + \rho_\mathrm{hol}^{(2)}\right) + \kappa^2\left(\rho_\mathrm{r}
+ \frac{1}{2}\dot{\phi}^2 + V(\phi)\right) \, .
\label{holographic_friedmann1}
\end{equation}
Clearly $\rho_\mathrm{hol}^{(1)}$ corresponds to the $R^2$ term
and thus dominates over the $\rho_\mathrm{hol}^{(2)}$ and
$\rho_\mathrm{r}$ term in the early epoch of the Universe.
Moreover, the axion field was frozen during the inflationary era
and thereby, we can safely neglect $\rho_{\phi}$ from
Eq.~(\ref{holographic_friedmann1}). Hence, in the early Universe
when the curvature is large, Eq.~(\ref{holographic_friedmann1})
behaves as,
\begin{equation}
3H^2 \simeq \kappa^2\rho_\mathrm{hol}^{(1)} \, ,
\label{holographic_friedmann2}
\end{equation}
which successfully produces an inflationary scenario. As the
Universe expands, the Hubble parameter decreases and from $H \ll
m_{\phi}$, the axion field starts to oscillate and thus
contributes its effect to the dynamics, along with the term
$\rho_\mathrm{r}$. Moreover, in the low curvature regime, as in
the case of present Universe, the energy density
$\rho_\mathrm{hol}^{(2)}$ dominates over the other terms of
Eq.~(\ref{holographic_friedmann1}). In view of these arguments,
after the inflationary scenario,
Eq.~(\ref{holographic_friedmann1}) becomes,
\begin{equation}
3H^2 \simeq \kappa^2\rho_\mathrm{hol}^{(2)} + \kappa^2\left(\rho_\mathrm{r} + \frac{1}{2}\dot{\phi}^2 + V(\phi)\right) \, ,
\label{holographic_friedmann3}
\end{equation}
where $\rho_\mathrm{r}$ and $\rho_{\phi}$ denote the radiation and
matter dominated epochs, respectively, while
$\rho_\mathrm{hol}^{(2)}$ denotes the holographic dark energy
density at late times. Therefore,
Eq.~(\ref{holographic_friedmann1}) is able to unify various
cosmological epochs of our Universe from a holographic point of
view.

Before closing this section, we would like to mention that apart
from the two aforementioned forms of $L_\mathrm{IR}$ (i.e., in
Eqs.~(\ref{axion1_hc_ph}) and (\ref{axion1_hc_fh})), an integral
form of $L_\mathrm{IR}$ for the axion-$F(R)$ model is also
calculated and given by,
\begin{align}
\frac{L_\mathrm{IR}}{c} = -\frac{1}{6\alpha\dot{H}^2a^6} \int dt a^6\dot{H}&\left[1-\gamma 6^{\delta-1}\left(2H^2+\dot{H}\right)^{\delta-2}
\left\{2H^2(2-\delta) + \frac{\dot{H}^2}{H^2}(1-\delta) - \frac{\ddot{H}}{H}\delta(1-\delta) + \dot{H}(4-7\delta+4\delta^2)\right\}
\right. \nonumber\\
& \left. -\frac{\kappa^2}{H^2}\left(\rho_\mathrm{r} + \rho_{\phi}\right)\right] \, .
\label{axion1_integral form1}
\end{align}
Inserting the above expression of cut-off into $H =
\frac{c}{L_\mathrm{IR}}$, one can reproduce the first Friedmann
Eq.~(\ref{axion1_eom}) for $f(R) = \frac{R^2}{M^2} - \gamma
R^{\delta}$. Therefore, the cut-off in Eq.~(\ref{axion1_integral
form1}) also provides a corresponding holographic model for the
axion-$F(R)$ action (\ref{axion1_action}).

%%%%%%%%%%%%%%%%%%%%%%%%%%%%%%%%%%%%%%%%%%%%%%%%%%%%%%%%%%%%%%%%%%%%%%%%%%%ch2

\subsubsection{Non-minimally coupled axion-$F(R)$ gravity model}

As an extension, we consider a second axion-$F(R)$ gravity model where the axion scalar field is non-minimally coupled
with the curvature, unlike to the previous model (\ref{axion1_action}) where the axion is minimally coupled with the gravity.
The action of the second axion-$F(R)$ model is
following \cite{Odintsov:2019evb}:
\begin{equation}
S = \int d^4x\sqrt{-g}\left[\frac{1}{2\kappa^2}\left(R + f(R,\phi)\right) - \frac{1}{2}\partial_{\mu}\phi\partial^{\mu}\phi
- V(\phi)\right] \, ,
\label{axion2_action}
\end{equation}
where $f(R,\phi)$ takes the form,
\begin{equation}
f(R,\phi) = \frac{R^2}{M^2} + h(\phi)R^{\delta} \, ,
\label{axion2_form of fR}
\end{equation}
where $\delta$ is a dimensionless parameter with values in the
interval $0 < \delta < 1$. It is evident that the axion field
couples with the curvature with the coupling function $h(\phi)$.
In the spatially flat FRW space-time, the first Friedmann equation
of the action (\ref{axion2_action}) becomes,
\begin{equation}
3H^2 = -\frac{f(R,\phi)}{2} + 3\left(H^2 + \dot{H}\right)\frac{\partial f}{\partial R}
 - 3H\frac{d}{dt}\left(\frac{\partial f}{\partial R}\right)
+ \frac{1}{2}\dot{\phi}^2 + V(\phi) \, ,
\label{axion2_eom}
\end{equation}
with the consideration that the axion field is homogeneous in
space. Similar to the previous model, the present model
(\ref{axion2_action}) is also able to unify various cosmological
epochs of our Universe like inflation, dark matter epoch and dark
energy epoch as described in \cite{Odintsov:2019evb}. During the
early epoch, the axion field in the action (\ref{axion2_action})
was frozen at its vacuum expectation value (vev) and due to the
consideration $V(\phi) \gg \frac{1}{\kappa^2}h(\phi)R^{\delta}$,
the axion field merely contributes a cosmological constant (due to
the presence of its potential) to the equation of motion. Again
for $m_{\phi} \sim \mathcal{O}(10^{-12})\, \mathrm{eV}$, the
cosmological constant coming from the axion vev can be neglected
compared to the $R^2$ term which is naturally dominant term in the
large curvature regime, as in the early Universe. Moreover due to
$0 < \delta < 1$, the $R^{\delta}$ term is sub-dominant in
comparison to the quadratic term and thus the dynamics of the
early Universe is controlled by the $R^2$ term which is known to
produce a viable inflationary era compatible with Planck
constraints. As the Universe expands, the Hubble parameter
decreases and from $H \sim m_{\phi}$, the axion begins its
dynamical evolution. Assuming slowly varying oscillating dynamics
for the axion, it can be shown that the energy density of the same
evolves as $\rho_{\phi} \sim a^{-3}$. Thus the axion field mimics
the behavior of the dark matter of the Universe. Finally during
the low curvature regime, i.e., in the present Universe, the term
$h(\phi)R^{\delta}$ starts to dominate and provides a dark energy
model. These arguments clearly indicate that the qualitative
nature of the models (\ref{axion1_action}) and
(\ref{axion2_action}) are more-or-less same. However it is worth
mentioning that the model (\ref{axion2_action}) predicts the
existence of a stiff matter era for the axion field at a
primordial pre-inflationary era, in which case the energy density
scales as $\rho_{\phi} \sim a^{-6}$, which is not predicted by the
model (\ref{axion1_action}). Actually the prediction of stiff
matter era from the model (\ref{axion2_action}) arises due to the
reason that the aforementioned condition $V(\phi) \gg
\frac{1}{\kappa^2}h(\phi)R^{\delta}$ should hold true during or
after the inflationary era. Therefore, in the pre-inflationary
epoch, one could have $V(\phi) \sim
\frac{1}{\kappa^2}h(\phi)R^{\delta}$ which in turn makes
$\rho_{\phi} \sim a^{-6}$ through the conservation equation of the
axion field.

Plugging the form of $f(R) = \frac{R^2}{M^2} + h(\phi)R^{\delta}$
into Eq.~(\ref{hc_ph_general}), we get the equivalent holographic
cut-off in terms of the particle horizon ($L_\mathrm{p}$) and its
derivatives as,
\begin{align}
\frac{c^2}{\left(L_\mathrm{IR}\right)^2}=&6\left(\frac{1}{L_\mathrm{p}} - \frac{\dot{L}_p}{L_\mathrm{p}}\right)
\left[\frac{2}{M^2} - \left(6\Omega_{\left(L_\mathrm{p} \right)}\right)^{\delta-2}h(\phi(a))\delta(1-\delta)\right]
\left[-4\frac{\dot{L}_p}{L_\mathrm{p}^3} + 6\frac{\dot{L}_p^2}{L_\mathrm{p}^3} - 2\frac{\dot{L}_p^3}{L_\mathrm{p}^3}
 - 3\frac{\ddot{L}_p}{L_\mathrm{p}^2} + \frac{\dot{L}_p\ddot{L}_p}{L_\mathrm{p}^2}
+ \frac{\dddot{L}_p}{L_\mathrm{p}}\right]\nonumber\\
&+ 6\Omega_{\left(L_\mathrm{p} \right)}\left[\frac{1}{L_\mathrm{p}^2} - \frac{\dot{L}_p}{L_\mathrm{p}^2}
+ \frac{\ddot{L}_p}{L_\mathrm{p}}\right]
\left[\frac{2}{M^2} + \left(6\Omega_{\left(L_\mathrm{p} \right)}\right)^{\delta-2}h(\phi(a))\delta\right]
 - 6\Omega_{\left(L_\mathrm{p} \right)}\left[\frac{2}{M^2} + \left(6\Omega_{\left(L_\mathrm{p} \right)}
\right)^{\delta-2}h(\phi(a))\right] \, ,
\label{axion2_hc_ph}
\end{align}
where $\Omega_{\left(L_\mathrm{p} \right)}$ is given after
Eq.~(\ref{axion1_hc_ph}) and $\phi=\phi(a)$ can be determined from
the conservation equation of the axion field. The holographic
cut-off for the model (\ref{axion2_action}) in terms of the
$L_\mathrm{f}$ and its derivatives is obtained from
Eq.~(\ref{hc_fh_general}) as follows,
\begin{align}
\frac{c^2}{\left(L_\mathrm{IR}\right)^2}=&6\left(-\frac{1}{L_\mathrm{f}} - \frac{\dot{L}_f}{L_\mathrm{f}}\right)
\left[\frac{2}{M^2} - \left(6\Omega_{\left(L_\mathrm{f} \right)}\right)^{\delta-2}h(\phi(a))\delta(1-\delta)\right]
\left[-4\frac{\dot{L}_f}{L_\mathrm{f}^3} - 6\frac{\dot{L}_f^2}{L_\mathrm{f}^3} - 2\frac{\dot{L}_f^3}{L_\mathrm{f}^3}
+ 3\frac{\ddot{L}_f}{L_\mathrm{f}^2} + \frac{\dot{L}_f\ddot{L}_f}{L_\mathrm{f}^2}
+ \frac{\dddot{L}_f}{L_\mathrm{f}}\right]\nonumber\\
&+ 6\Omega_{\left(L_\mathrm{f} \right)}\left[\frac{1}{L_\mathrm{f}^2} + \frac{\dot{L}_f}{L_\mathrm{f}^2}
+ \frac{\ddot{L}_f}{L_\mathrm{f}}\right]
\left[\frac{2}{M^2} + \left(6\Omega_{\left(L_\mathrm{f} \right)}\right)^{\delta-2}h(\phi(a))\delta\right]
 - 6\Omega_{\left(L_\mathrm{f} \right)}\left[\frac{2}{M^2} + \left(6\Omega_{\left(L_\mathrm{f} \right)}
\right)^{\delta-2}h(\phi(a))\right] \, ,
\label{axion2_hc_fh}
\end{align}
for $\Omega_{\left(L_\mathrm{f} \right)}$, see the expression just
after Eq.~(\ref{axion1_hc_fh}). Clearly the axion-$F(R)$ model
(\ref{axion2_action}) is equivalent to the holographic model with
the cut-offs determined in the above two expressions. The
corresponding holographic Friedmann equation takes the form:
\begin{equation}
3H^2 = \frac{3c^2}{\left(L_\mathrm{IR}\right)^2} + \kappa^2\left(\frac{1}{2}\dot{\phi}^2 + V(\phi)\right) \, .
\label{axion2_holographic_friedmann}
\end{equation}
The cut-offs determined in Eqs.~(\ref{axion2_hc_ph}) and
(\ref{axion2_hc_fh}) are decomposed as,
\begin{equation}
\frac{1}{\left(L_\mathrm{IR}\right)^2} = \frac{1}{\left(L^{(1)}_\mathrm{IR}\right)^2}
+ \frac{1}{\left(L^{(2)}_\mathrm{IR}\right)^2} \, , \nonumber
\end{equation}
and the above expression can be rewritten as,
\begin{equation}
\rho_\mathrm{hol} = \rho_\mathrm{hol}^{(1)} + \rho_\mathrm{hol}^{(2)} \, ,
\label{axion2_decomposition}
\end{equation}
where, in spirit of Eq.~(\ref{basic}), $\rho_\mathrm{hol}^{(i)} =
\frac{3c^2}{\kappa^2\left(L^{(i)}_\mathrm{IR}\right)^2}$.
Furthermore  $\rho_\mathrm{hol}^{(1)}$ and
$\rho_\mathrm{hol}^{(2)}$ are given by,
\begin{align}
&\rho_\mathrm{hol}^{(1)}=\frac{3c^2}{\kappa^2\left(L^{(1)}_\mathrm{IR}\right)^2}\nonumber\\
=&\frac{3}{\kappa^2}\bigg\{\frac{12}{M^2}\left(\frac{1}{L_\mathrm{p}} - \frac{\dot{L}_p}{L_\mathrm{p}}\right)
\left[-4\frac{\dot{L}_p}{L_\mathrm{p}^3} + 6\frac{\dot{L}_p^2}{L_\mathrm{p}^3} - 2\frac{\dot{L}_p^3}{L_\mathrm{p}^3}
 - 3\frac{\ddot{L}_p}{L_\mathrm{p}^2} + \frac{\dot{L}_p\ddot{L}_p}{L_\mathrm{p}^2}
+ \frac{\dddot{L}_p}{L_\mathrm{p}}\right] + \frac{12}{M^2}\Omega_{\left(L_\mathrm{p} \right)}\left[\frac{1}{L_\mathrm{p}^2}
 - \frac{\dot{L}_p}{L_\mathrm{p}^2} + \frac{\ddot{L}_p}{L_\mathrm{p}}\right]
 - \frac{12}{M^2}\Omega_{\left(L_\mathrm{p} \right)}\bigg\} \nonumber\\
=&\frac{3}{\kappa^2}\bigg\{\frac{12}{M^2}\left(-\frac{1}{L_\mathrm{f}} - \frac{\dot{L}_f}{L_\mathrm{f}}\right)
\left[-4\frac{\dot{L}_f}{L_\mathrm{f}^3} - 6\frac{\dot{L}_f^2}{L_\mathrm{f}^3}
 - 2\frac{\dot{L}_f^3}{L_\mathrm{f}^3} + 3\frac{\ddot{L}_f}{L_\mathrm{f}^2} + \frac{\dot{L}_f\ddot{L}_f}{L_\mathrm{f}^2}
+ \frac{\dddot{L}_f}{L_\mathrm{f}}\right]
+ \frac{12}{M^2}\Omega_{\left(L_\mathrm{f} \right)}\left[\frac{1}{L_\mathrm{f}^2}
+ \frac{\dot{L}_f}{L_\mathrm{f}^2} + \frac{\ddot{L}_f}{L_\mathrm{f}}\right]
- \frac{12}{M^2}\Omega_{\left(L_\mathrm{f} \right)}\bigg\}\, , \nonumber\\
\label{axion2_cut-off_early}
\end{align}
and
\begin{align}
\rho_\mathrm{hol}^{(2)} =& \frac{3c^2}{\kappa^2\left(L^{(2)}_\mathrm{IR}\right)^2} \nonumber \\
=&\frac{3}{\kappa^2}\bigg\{-6\left(6\Omega_{\left(L_\mathrm{p} \right)}\right)^{\delta-2}h(\phi(a))\delta(1-\delta)\left(\frac{1}{L_\mathrm{p}}
 - \frac{\dot{L}_p}{L_\mathrm{p}}\right)
\left[-4\frac{\dot{L}_p}{L_\mathrm{p}^3} + 6\frac{\dot{L}_p^2}{L_\mathrm{p}^3}
 - 2\frac{\dot{L}_p^3}{L_\mathrm{p}^3} - 3\frac{\ddot{L}_p}{L_\mathrm{p}^2} + \frac{\dot{L}_p\ddot{L}_p}{L_\mathrm{p}^2}
+ \frac{\dddot{L}_p}{L_\mathrm{p}}\right]\nonumber\\
&+\left(6\Omega_{\left(L_\mathrm{p} \right)}\right)^{\delta-1}h(\phi(a))\delta\left[\frac{1}{L_\mathrm{p}^2}
 - \frac{\dot{L}_p}{L_\mathrm{p}^2} + \frac{\ddot{L}_p}{L_\mathrm{p}}\right]
 - h(\phi(a))\left(6\Omega_{\left(L_\mathrm{p} \right)}\right)^{\delta-1}\bigg\} \nonumber\\
=&\frac{3}{\kappa^2}\bigg\{-3\left(6\Omega_{\left(L_\mathrm{f} \right)}\right)^{\delta-2}h(\phi(a))\delta(1-\delta)\left(-\frac{1}{L_\mathrm{f}}
 - \frac{\dot{L}_f}{L_\mathrm{f}}\right)
\left[-4\frac{\dot{L}_f}{L_\mathrm{f}^3} - 6\frac{\dot{L}_f^2}{L_\mathrm{f}^3}
 - 2\frac{\dot{L}_f^3}{L_\mathrm{f}^3} + 3\frac{\ddot{L}_f}{L_\mathrm{f}^2} + \frac{\dot{L}_f\ddot{L}_f}{L_\mathrm{f}^2}
+ \frac{\dddot{L}_f}{L_\mathrm{f}}\right]\nonumber\\
&+\left(6\Omega_{\left(L_\mathrm{f} \right)}\right)^{\delta-1}h(\phi(a))\delta\left[\frac{1}{L_\mathrm{f}^2}
+ \frac{\dot{L}_f}{L_\mathrm{f}^2} + \frac{\ddot{L}_f}{L_\mathrm{f}}\right]
 - h(\phi(a))\left(6\Omega_{\left(L_\mathrm{f} \right)}\right)^{\delta-1}\bigg\} \, ,
\label{axion2_cut-off_late}
\end{align}
respectively. With such decomposition of $L_\mathrm{IR}$,
Eq.~(\ref{axion2_holographic_friedmann}) can be rewritten as,
\begin{equation}
3H^2 = \kappa^2\left(\rho_\mathrm{hol}^{(1)} + \rho_\mathrm{hol}^{(2)}\right)
+ \kappa^2\left(\frac{1}{2}\dot{\phi}^2 + V(\phi)\right) \, .
\label{axion2_holographic_friedmann1}
\end{equation}
As earlier, $\rho_\mathrm{hol}^{(1)}$ corresponds to the $R^2$
term and due to the arguments demonstrated just after
Eq.~(\ref{axion2_eom}), the holographic
Eq.~(\ref{axion2_holographic_friedmann1}) behaves during the early
Universe as,
\begin{equation}
3H^2 \simeq \kappa^2\rho_\mathrm{hol}^{(1)}\, ,
\label{axion2_holographic_friedmann2}
\end{equation}
which describes an inflationary scenario with good agreement in
terms of the Planck observations. On other hand, after the
inflationary scenario, the axion field starts to contribute and
also in the present Universe the $h(\phi)R^{\delta}$ term
dominates over the quadratic curvature. As a result, after
inflation, Eq.~(\ref{axion2_holographic_friedmann1}) becomes,
\begin{equation}
3H^2 \simeq \kappa^2\rho_\mathrm{hol}^{(2)} + \kappa^2\left(\frac{1}{2}\dot{\phi}^2 + V(\phi)\right) \, ,
\label{axion2_holographic_friedmann3}
\end{equation}
where $\rho_{\phi}$ denotes the matter dominated epoch, while
$\rho_\mathrm{hol}^{(2)}$ stands for the holographic dark energy
density during late time of the Universe. Therefore,
Eq.~(\ref{axion2_holographic_friedmann1}) is able to unify various
cosmological epochs of our Universe like inflation, dark matter
and dark energy epochs respectively, from a holographic point of
view.

Similar to an earlier case, here we also determine the integral
form of $L_\mathrm{IR}$ for the non-minimally coupled axion-$F(R)$
model, which is,
\begin{align}
\frac{L_\mathrm{IR}}{c} = -\frac{1}{6\alpha\dot{H}^2a^6} \int dt a^6\dot{H}& \left[1+h(\phi(a))6^{\delta-1}\left(2H^2+\dot{H}\right)^{\delta-2}
\right. \nonumber\\
& \left. \times \left\{2H^2(2-\delta) + \frac{\dot{H}^2}{H^2}(1-\delta) - \frac{\ddot{H}}{H}\delta(1-\delta)
+ \dot{H}(4-7\delta+4\delta^2)\right\}
 - \frac{\kappa^2}{H^2}\left(\rho_\mathrm{r} + \rho_{\phi}\right)\right] \, .
\label{axion2_integral form1}
\end{align}
Inserting the above expression of cut-off into $H =
\frac{c}{L_\mathrm{IR}}$, one can reproduce Eq.~(\ref{axion2_eom})
for $f(R,\phi) = \frac{R^2}{M^2} + h(\phi)R^{\delta}$. Therefore,
the axion-$F(R)$ model (\ref{axion2_action}) can also be mapped to
a holographic model with the cut-off determined in
Eq.~(\ref{axion2_integral form1}).

%%%%%%%%%%%%%%%%%%%%%%%%%%%%%%%%%%%%%%%%%%%%%%%%%%%%%%%%%%%%%%%%%%%%%%%%%%%%%%%%%%%%%%%%%%%%%%%%%%%%%%%%%%%%%%%%%%%%%%%ch1

\section{Holographic correspondence of $f(\mathcal{G})$ gravity}\label{sec_general_f(G)}

We now establish the holographic correspondence and consequently
determine the holographic cut-off for $f(\mathcal{G})$ gravity
whose action is given by (see \cite{Nojiri:2010wj,Li:2007jm} for
different aspects of $f(\mathcal{G})$ gravity),
\begin{equation}
S=\int d^4x\sqrt{-g} \left[\frac{1}{2\kappa^2}R + f(\mathcal{G}) + \mathcal{L}_\mathrm{mat}\right] \, ,
\label{action_f(G)}
\end{equation}
where $\mathcal{G} = R^2 - 4R_{\mu\nu}R^{\mu\nu} + R_{\mu\nu\alpha\beta}R^{\mu\nu\alpha\beta}$ 
is the Gauss-Bonnet invariant which,
in the FRW space-time, takes the form as,
\begin{equation}
\mathcal{G} = 24H^2\left[H^2 + \dot{H}\right] \, .
\label{G}
\end{equation}
With this expression of $\mathcal{G}$, the first Friedmann equation can be written as,
\begin{equation}
0=-\frac{3}{\kappa^2}H^2 - f(\mathcal{G}) + \mathcal{G}f'(\mathcal{G}) - 24 \dot{\mathcal{G}}f''(\mathcal{G}) H^3
+ \rho_\mathrm{mat}\, .
\label{eom_f(G)}
\end{equation}
It is interesting to note that for $f(\mathcal{G}) = \mathcal{G}$, Eq.(\ref{eom_f(G)}) reduces to the standard Friedmann 
equation for Einstein gravity. This is expected because the Gauss-Bonnet term in 3+1 dimensional spacetime becomes a topological surface term and 
therefore vanishes identically. However the functional form other than $f(\mathcal{G}) = \mathcal{G}$ indeed contributes in the equation of motion as 
reflected from the above equation. Comparing Eq.~(\ref{eom_f(G)}) with the holographic Friedmann
equation $H^2 = \frac{1}{\left(L_\mathrm{IR}\right)^2}$, we can
immediately conclude that $f(\mathcal{G})$ gravity (without/with
matter fields) has an equivalent holographic correspondence with
the holographic cut-off is given by,
\begin{equation}
\frac{3c^2}{\left(L_\mathrm{IR}\right)^2} = \kappa^2\left[- f(\mathcal{G})
+ \mathcal{G}f'(\mathcal{G}) - 24 \dot{\mathcal{G}}f''(\mathcal{G}) H^3\right] \, ,
\label{hc_general_f(G)}
\end{equation}
and as a consequence, Eq.~(\ref{eom_f(G)}) is rewritten as $3H^2 =
\frac{3}{\left(L_\mathrm{IR}\right)^2} +
\kappa^2\rho_\mathrm{mat}$. Thereby, any arbitrary
$f(\mathcal{G})$ gravity can be mapped to an equivalent
holographic model with the cut-off given by
Eq.~(\ref{hc_general_f(G)}). In a similar way as in $F(R)$
gravity, here we also determine $L_\mathrm{IR}$ in two different
ways - namely in terms of the particle horizon ($L_\mathrm{p}$)
and its derivatives or in terms of the future horizon
($L_\mathrm{f}$) and its derivatives. Recall that the Hubble
parameter in terms of $L_\mathrm{p}$ or $L_\mathrm{f}$ can be
expressed as $H(L_\mathrm{p},\dot{L}_p) =
\frac{\dot{L}_p}{L_\mathrm{p}} - \frac{1}{L_\mathrm{p}}$ or
$H(L_\mathrm{f},\dot{L}_f) = \frac{\dot{L}_f}{L_\mathrm{f}} +
\frac{1}{L_\mathrm{f}}$ respectively. These considerations lead to
the Gauss-Bonnet invariant as,
\begin{align}
\mathcal{G}^{\left(L_\mathrm{p} \right)}=&24\left(\frac{\dot{L}_p}{L_\mathrm{p}}
 - \frac{1}{L_\mathrm{p}}\right)^2\left(\frac{\ddot{L}_p}{L_\mathrm{p}} - \frac{\dot{L}_p}{L_\mathrm{p}^2}
+ \frac{1}{L_\mathrm{p}^2}\right) \, ,
\label{G_ph}\\
\mathcal{G}^{\left(L_\mathrm{f} \right)}=&24\left(\frac{\dot{L}_f}{L_\mathrm{f}}
+ \frac{1}{L_\mathrm{f}}\right)^2\left(\frac{\ddot{L}_f}{L_\mathrm{f}} + \frac{\dot{L}_f}{L_\mathrm{f}^2}
+ \frac{1}{L_\mathrm{f}^2}\right)\label{G_fh} \, .
\end{align}
Using the above expressions, the holographic cut-off in terms of $L_\mathrm{p}$ and its derivatives can be determined as,
\begin{equation}
\frac{3c^2}{\left(L_\mathrm{IR}\right)^2} = f \left(\mathcal{G}^{\left(L_\mathrm{p} \right)} \right)
 - \mathcal{G}^{\left(L_\mathrm{p} \right)}f' \left(\mathcal{G}^{\left(L_\mathrm{p} \right)} \right)
+ 24\frac{d\mathcal{G}^{\left(L_\mathrm{p} \right)}}{dt}
f'' \left(\mathcal{G}^{\left(L_\mathrm{p} \right)}\right)\left(\frac{\dot{L}_p}{L_\mathrm{p}}
 - \frac{1}{L_\mathrm{p}}\right)^3 \, .
\label{hc_ph_generalf(G)}
\end{equation}
Similarly, $L_\mathrm{IR} =
L_\mathrm{IR}\left(L_\mathrm{f},\dot{L}_\mathrm{f},\ddot{L}_\mathrm{f}, 
\ \mbox{higher derivatives of}\ L_\mathrm{f} \right)$ takes the
following form,
\begin{equation}
 \frac{3c^2}{\left(L_\mathrm{IR}\right)^2} = f \left(\mathcal{G}^{\left(L_\mathrm{f} \right)} \right)
- \mathcal{G}^{\left(L_\mathrm{f} \right)}f' \left(\mathcal{G}^{\left(L_\mathrm{f} \right)} \right)
+ 24\frac{d\mathcal{G}^{\left(L_\mathrm{f} \right)}}{dt}
f'' \left(\mathcal{G}^{\left(L_\mathrm{f} \right)}\right) \left(\frac{\dot{L}_f}{L_\mathrm{f}}
+ \frac{1}{L_\mathrm{f}}\right)^3 \, .
\label{hc_fh_generalf(G)}
\end{equation}
Eqs.~(\ref{hc_ph_generalf(G)}) and (\ref{hc_fh_generalf(G)}) are
the key equations that will determine the equivalent holographic
cut-off for any arbitrary $f(\mathcal{G})$ gravity. As an example,
we may consider a simple model like
\begin{equation}
f(\mathcal{G}) = f_0\mathcal{G}^m \, ,
\label{example1_f(G)}
\end{equation}
where $f_0$ and $m$ are dimensionless parameters
\cite{Cognola:2006eg}. Plugging back this explicit
form of $f(\mathcal{G})$ into Eq.~(\ref{hc_ph_generalf(G)}), we
get,
\begin{align}
\frac{c^2}{\left(L_\mathrm{IR}\right)^2}=&\frac{f_0(1-m)}{3\left(1 - \dot{L}_p
+ L_\mathrm{p}\ddot{L}_p\right)^2}\left[\frac{24 \left(1 - \dot{L}_p\right)^2
\left(1 - \dot{L}_p + L_\mathrm{p}\ddot{L}_p\right)}{L_\mathrm{p}^4}\right]^m
\left[1 - 4m\dot{L}_p^3 + (2-3m)L_\mathrm{p}\ddot{L}_p \right. \nonumber\\
& \left. + \dot{L}_p^2\left(1 + 8m + 3mL_\mathrm{p}\ddot{L}_p\right)
+ L_\mathrm{p}^2\left((1-2m)\ddot{L}_p^2 + m\dddot{L}_p\right)
 - \dot{L}_p\left(2 + 4m + 2L_\mathrm{p}\ddot{L}_p + mL_\mathrm{p}^2\dddot{L}_p\right)\right] \, .
\label{example1_f(G)_hc_ph}
\end{align}
Moreover Eq.~(\ref{hc_fh_generalf(G)}) leads to the holographic
cut-off for $f(\mathcal{G})=f_0\mathcal{G}^m$ in terms of the
future horizon as,
\begin{align}
\frac{c^2}{\left(L_\mathrm{IR}\right)^2}=&\frac{f_0(1-m)}{3\left(1 + \dot{L}_f
+ L_\mathrm{f}\ddot{L}_f\right)^2}\left[\frac{24 \left(1 + \dot{L}_f\right)^2
\left(1 + \dot{L}_f + L_\mathrm{f}\ddot{L}_f\right)}{L_\mathrm{p}^4}\right]^m
\left[1 + 4m\dot{L}_f^3 + (2-3m)L_\mathrm{f}\ddot{L}_f \right.\nonumber\\
& \left. +\dot{L}_f^2\left(1 + 8m + 3mL_\mathrm{f}\ddot{L}_f\right)
+ L_\mathrm{f}^2\left((1-2m)\ddot{L}_f^2 - m\dddot{L}_f\right)
+ \dot{L}_f\left(2 + 4m + 2L_\mathrm{f}\ddot{L}_f - mL_\mathrm{f}^2\dddot{L}_f\right)\right] \, .
\label{example1_f(G)_hc_fh}
\end{align}
Clearly the above two holographic cut-offs can mimic the
cosmological field equations and thus we established the
holographic correspondence for the $f(\mathcal{G}) =
f_0\mathcal{G}^m$ model.

It may be observed from Eq.~(\ref{example1_f(G)}) that for
$m>1/2$, the term $f(\mathcal{G})\sim\mathcal{G}^m$ dominates over
the Einstein and the matter term(s) in the large curvature regime,
while for $m<1/2$, the Gauss-Bonnet function $f(\mathcal{G})$
becomes the dominating one in the low curvature regime. Here we
consider a case in which the contributions from the Einstein and
matter terms can be neglected compared to
$f(\mathcal{G})\sim\mathcal{G}^m$. In such situation, the scale
factor evolves as $a(t) = a_0 t^{h_0}$ where the exponent $h_0$ is
given by: $h_0 = 1 - 4m$. Such evolution of the scale factor
immediately leads to the effective EoS parameter as,
\begin{equation}
w_\mathrm{eff} = -1 + \frac{2}{3h_0} = -1 + \frac{2}{3(1-4m)}\, ,
\label{EoS}
\end{equation}
where in the last line, we used the expression for $h_0$.
Therefore, if $m > 0$, the Universe is accelerating
($w_\mathrm{eff} < -1/3$), and if $m > 1/4$, the Universe is in a
phantom phase ($w_\mathrm{eff} < -1$). Hence, depending on the
parameter $m$, the model $f(\mathcal{G}) = f_0\mathcal{G}^m$ can
act as inflationary or as late-time accelerating model. However,
this model does not give an unified scenario of inflation and late
time dark energy epoch of our Universe. Keeping this in mind, we
will consider a different form of $f(\mathcal{G})$, in the next
subsection, that is able to describe inflation and the late-time
acceleration of the Universe in a unified manner.

\subsection{Unification of holographic inflation with holographic dark energy}\label{sec_unification_f(G)}

In spirit of the power law model and the above discussions, we consider the following model \cite{Nojiri:2010wj},
\begin{equation}
f(\mathcal{G}) = f_1\mathcal{G}^{\beta_1} + f_2\mathcal{G}^{\beta_2} \, ,
\label{example2_f(G)}
\end{equation}
where the exponents are assumed to take values in the intervals,
\begin{equation}
\beta_1 > \frac{1}{2}\, , \quad \frac{1}{4} < \beta_2 < \frac{1}{2} \, .
\label{exponent}
\end{equation}
Thus in the large curvature regime, as in the early Universe, the
first term dominates compared to the second term and the Einstein
term, which in turn leads to the effective EoS parameter as
$w^{(1)}_\mathrm{eff} = -1 + \frac{2}{3(1 - 4\beta_1)}$ (using
Eq.~(\ref{EoS})). Due to $\beta_1 > 1/2$,
\begin{equation}
 -\frac{5}{3} < w^{(1)}_\mathrm{eff} < -1 \, .
\label{EoS1}
\end{equation}
On the other hand, when the curvature is small, as is the case in
the late Universe, the second term in (\ref{example2_f(G)})
dominates compared with the first term and the Einstein term and
yields,
\begin{equation}
w^{(1)}_\mathrm{eff} = -1 + \frac{2}{3(1 - 4\beta_2)} < -\frac{5}{3} \, .
\label{EoS2}
\end{equation}
Therefore, the theoretical framework (\ref{example2_f(G)})
produces a model that describes the unified scenario of inflation
and dark energy epochs of our Universe. By inserting the explicit
form of $f(\mathcal{G}) = f_1\mathcal{G}^{\beta_1} +
f_2\mathcal{G}^{\beta_2}$ into Eq.~(\ref{hc_ph_generalf(G)}), we
get the holographic cut-off in terms of $L_\mathrm{p}$ and its
derivatives as,
\begin{align}
\frac{c^2}{\left(L_\mathrm{IR}\right)^2}=&\frac{f_1 }{3}
\left[\frac{24 \left(1 - \dot{L}_p\right)^2\left(1 - \dot{L}_p + L_\mathrm{p}\ddot{L}_p\right)}{L_\mathrm{p}^4}\right]^{\beta_1}
\left[1 - \beta_1 - \beta_1(1-\beta_1)\frac{X\left(L_\mathrm{p},\dot{L}_p,\ddot{L}_p,\dddot{L}_p\right)}
{\left(1 - \dot{L}_p + L_\mathrm{p}\ddot{L}_p\right)^2}\right]\nonumber\\
+&\frac{f_2}{3}
\left[\frac{24\left(1 - \dot{L}_p\right)^2\left(1 - \dot{L}_p + L_\mathrm{p}\ddot{L}_p\right)}{L_\mathrm{p}^4}\right]^{\beta_2}
\left[1 - \beta_2 - \beta_2(1-\beta_2)\frac{X\left(L_\mathrm{p},\dot{L}_p,\ddot{L}_p,\dddot{L}_p\right)}
{\left(1 - \dot{L}_p + L_\mathrm{p}\ddot{L}_p\right)^2}\right] \, .
\label{example2_f(G)_hc_ph}
\end{align}
Eq.~(\ref{hc_fh_generalf(G)}) leads to the $L_\mathrm{IR}$ as
function of the future horizon and its derivatives as,
\begin{align}
\frac{c^2}{\left(L_\mathrm{IR}\right)^2}=&\frac{f_1}{3}
\left[\frac{24 \left(1 + \dot{L}_f\right)^2\left(1 + \dot{L}_f + L_\mathrm{f}\ddot{L}_f\right)}{L_\mathrm{f}^4}\right]^{\beta_1}
\left[1 - \beta_1 - \beta_1(1-\beta_1)\frac{\Upsilon\left(L_\mathrm{f},\dot{L}_f,\ddot{L}_f,\dddot{L}_f\right)}
{\left(1 + \dot{L}_f + L_\mathrm{f}\ddot{L}_f\right)^2}\right]
\nonumber\\
&+\frac{f_2}{3}
\left[\frac{24 \left(1 + \dot{L}_f\right)^2\left(1 + \dot{L}_f + L_\mathrm{f}\ddot{L}_f\right)}{L_\mathrm{f}^4}\right]^{\beta_2}
\left[1 - \beta_2 - \beta_2(1-\beta_2)\frac{\Upsilon\left(L_\mathrm{f},\dot{L}_f,\ddot{L}_f,\dddot{L}_f\right)}
{\left(1 + \dot{L}_f + L_\mathrm{p}\ddot{L}_p\right)^2}\right] \, ,
\label{example2_f(G)_hc_fh}
\end{align}
where the functions $X$ and $\Upsilon$ are given by,
\begin{equation}
X\left(L_\mathrm{p},\dot{L}_p,\ddot{L}_p,\dddot{L}_p\right)=4\dot{L}_p - 8\dot{L}_p^2 + 4\dot{L}_p^3
 - 3L_\mathrm{p}\dot{L}_p^2\ddot{L}_p + L_\mathrm{p}^2\dot{L}_p\dddot{L}_p
+ L_\mathrm{p}\left(3\ddot{L}_p + 2L_\mathrm{p}\ddot{L}_p^2 - L_\mathrm{p}\dddot{L}_p\right) \, , \nonumber
\end{equation}
and
\begin{equation}
\Upsilon\left(L_\mathrm{f},\dot{L}_f,\ddot{L}_f,\dddot{L}_f\right)=-4\dot{L}_f - 8\dot{L}_f^2
 - 4\dot{L}_f^3 - 3L_\mathrm{f}\dot{L}_f^2\ddot{L}_f
+ L_\mathrm{f}^2\dot{L}_f\dddot{L}_f
+ L_\mathrm{f}\left(3\ddot{L}_f + 2L_\mathrm{f}\ddot{L}_f^2 + L_\mathrm{f}\dddot{L}_f\right) \, ,
\label{functions}
\end{equation}
respectively. The cut-offs determined in
Eq.~(\ref{example2_f(G)_hc_ph}) or Eq.~(\ref{example2_f(G)_hc_fh})
along with the expression $H^2 =
\frac{1}{\left(L_\mathrm{IR}\right)^2}$ can reconstruct the
cosmological field equations and thus provide an equivalent
holographic scenario for the considered $f(\mathcal{G})$ model
(\ref{example2_f(G)}). The cut-offs determined in
Eqs.~(\ref{example2_f(G)_hc_ph}) and (\ref{example2_f(G)_hc_fh})
can be decomposed as,
\begin{equation}
\frac{1}{\left(L_\mathrm{IR}\right)^2} = \frac{1}{\left(L^{(1)}_\mathrm{IR}\right)^2}
+ \frac{1}{\left(L^{(2)}_\mathrm{IR}\right)^2} \, ,\nonumber
\end{equation}
or the above expression can be rewritten as,
\begin{equation}
\rho_\mathrm{hol} = \rho_\mathrm{hol}^{(1)} + \rho_\mathrm{hol}^{(2)} \, ,
\label{decomposition}
\end{equation}
where $\rho_\mathrm{hol}^{(i)} = \frac{3c^2}{\kappa^2\left(L^{(i)}_\mathrm{IR}\right)^2}$.
Furthermore  $\rho_\mathrm{hol}^{(1)}$ and $\rho_\mathrm{hol}^{(2)}$ are given by,
\begin{align}
\rho_\mathrm{hol}^{(1)}=&\frac{f_1}{\kappa^2}
\left[\frac{24 \left(1 - \dot{L}_p\right)^2\left(1 - \dot{L}_p + L_\mathrm{p}\ddot{L}_p\right)}{L_\mathrm{p}^4}\right]^{\beta_1}
\left[1 - \beta_1 - \beta_1(1-\beta_1)\frac{X\left(L_\mathrm{p},\dot{L}_p,\ddot{L}_p,\dddot{L}_p\right)}
{\left(1 - \dot{L}_p + L_\mathrm{p}\ddot{L}_p\right)^2}\right] \nonumber\\
=&\frac{f_1}{\kappa^2}
\left[\frac{24 \left(1 + \dot{L}_f\right)^2\left(1 + \dot{L}_f + L_\mathrm{f}\ddot{L}_f\right)}{L_\mathrm{f}^4}\right]^{\beta_1}
\left[1 - \beta_1 - \beta_1(1-\beta_1)\frac{\Upsilon\left(L_\mathrm{f},\dot{L}_f,\ddot{L}_f,\dddot{L}_f\right)}
{\left(1 + \dot{L}_f + L_\mathrm{f}\ddot{L}_f\right)^2}\right] \, ,
\label{cut-off_early}
\end{align}
and
\begin{align}
\rho_\mathrm{hol}^{(2)}=&\frac{f_2}{\kappa^2}
\left[\frac{24 \left(1 - \dot{L}_p\right)^2\left(1 - \dot{L}_p + L_\mathrm{p}\ddot{L}_p\right)}{L_\mathrm{p}^4}\right]^{\beta_2}
\left[1 - \beta_2 - \beta_2(1-\beta_2)\frac{X\left(L_\mathrm{p},\dot{L}_p,\ddot{L}_p,\dddot{L}_p\right)}
{\left(1 - \dot{L}_p + L_\mathrm{p}\ddot{L}_p\right)^2}\right] \nonumber\\
=&\frac{f_2}{\kappa^2}
\left[\frac{24 \left(1 + \dot{L}_f\right)^2\left(1 + \dot{L}_f + L_\mathrm{f}\ddot{L}_f\right)}{L_\mathrm{f}^4}\right]^{\beta_2}
\left[1 - \beta_2 - \beta_2(1-\beta_2)\frac{\Upsilon\left(L_\mathrm{f},\dot{L}_f,\ddot{L}_f,\dddot{L}_f\right)}{\left(1 + \dot{L}_f
+ L_\mathrm{p}\ddot{L}_p\right)^2}\right] \, ,
\label{cut-off_late}
\end{align}
respectively. With such decomposition, the holographic Friedmann equation turns out to be,
\begin{equation}
3H^2 = \kappa^2\left(\rho_\mathrm{hol}^{(1)} + \rho_\mathrm{hol}^{(2)}\right) \, .
\label{holographic_friedmann_fG}
\end{equation}
Clearly $\rho_\mathrm{hol}^{(1)}$ corresponds to
$f_1\mathcal{G}^{\beta_1}$ and thus dominates over
$\rho_\mathrm{hol}^{(2)}$ in the large curvature regime, while in
the low curvature regime $\rho_\mathrm{hol}^{(2)}$ is the dominant
compared to the other one. Therefore, during early Universe when
the curvature is large, Eq.~(\ref{holographic_friedmann_fG}) can
be approximated as $3H^2 \simeq \kappa^2\rho_\mathrm{hol}^{(1)}$
which produces an inflationary scenario. On other hand due to low
curvature in the present Universe,
Eq.~(\ref{holographic_friedmann_fG}) goes as $3H^2 \simeq
\kappa^2\rho_\mathrm{hol}^{(2)}$ which provides a holographic dark
energy model during late-time. Thereby, the holographic Friedmann
Eq.~(\ref{holographic_friedmann_fG}) (see
Eqs.~(\ref{cut-off_early}) and (\ref{cut-off_late}) for the
expressions of $\rho_\mathrm{hol}^{(i)}$) is able to describe
inflation and dark energy epochs of the Universe in a unified way.

\section{Holographic correspondence of $F(T)$ gravity}\label{sec_general_F(T)}

We extend our discussion of holographic correspondence to the generalized teleparallel cosmology, i.e., $F(T)$ cosmology
\cite{Cai:2015emx,Krssak:2015oua}.
The teleparallel gravity (TEGR) is described by the Weitzenbock connection which
is determined by two dynamical variables, namely the tetrads and the
spin connection. Moreover unlike to the connection in Einstein's General Relativity, the Weitzenbock connection comes as
a curvature-free quantity. Recall, in the current work, we consider the
spatially flat FRW metric, i.e.,
\begin{equation}
ds^2 = -dt^2 + a^2(t)\left[dx^2 + dy^2 + dz^2\right] \, ,
\label{t1}
\end{equation}
with $a(t)$ being the scale factor of the Universe. Generally the tetrad for this metric is considered as,
\begin{equation}
e^a_{\mu} = \mathrm{diag} \left(1,a(t),a(t),a(t)\right) \, .
\label{t2}
\end{equation}
This form of the FRW metric is very advantageous because its spin
connection vanishes \cite{Krssak:2015oua}, and so no extra
contribution is needed in the $F(T)$ field equations. For the
aforementioned tetrad in Eq.~(\ref{t2}), the torsion scalar turns
out to be $T = -6H^2$, with $H$ being the Hubble parameter.

%%%%%%%%%%%%%%%%%%%
%%%%%%%%%%%%%%%%%%%

Following the same reasoning as $F(R)$ gravity, the action of TEGR can be generalized to $F(T)$ gravity, in particular
\begin{equation}
S = \frac{1}{2\kappa^2} \int d^4x |e|~F(T) \, ,
\label{t3}
\end{equation}
where $e = \det~e^a_{\mu}$ (We use $e = \det~e^a_{\mu}$ in italics
and $\e$ for Napier's constant $\e=2.718281828\cdots$) and $F(T)$
is an analytic function of $T$. Variation of the action with
respect to the tetrad in the FRW spacetime yields the following
equation:
\begin{equation}
H^2 = -\frac{\left(F(T) - T\right)}{6} - 2H^2~\frac{dF}{dT} \, .
\label{t4}
\end{equation}
The above differential equation can be mapped to the holographic
Friedmann equation $H = \frac{1}{L_\mathrm{IR}}$, with the holographic
cut-off being,
\begin{equation}
\frac{c^2}{\left(L_\mathrm{IR}\right)^2} = -\frac{\left(F(T) - T\right)}{6} - 2H^2~\frac{dF}{dT} \, .
\label{hc_general_F(T)}
\end{equation}
As similar to earlier gravity theories, here we also determine the cut-off in terms of the particle horizon and the future horizon.
For this purpose, what we need is the following expression:
\begin{equation}
T^{\left(L_\mathrm{p} \right)} 
= -6\left[\frac{\dot{L}_\mathrm{p}}{L_\mathrm{p}} - \frac{1}{L_\mathrm{p}}\right]^2 \, , \quad
T^{\left(L_\mathrm{f} \right)} 
= -6\left[\frac{\dot{L}_\mathrm{f}}{L_\mathrm{f}} + \frac{1}{L_\mathrm{f}}\right]^2 \, . \nonumber
\end{equation}
The above expressions along with Eq.~(\ref{hc_general_F(T)})
immediately lead to the $L_\mathrm{IR} = L_\mathrm{IR}
\left(L_\mathrm{p} , \dot{L}_\mathrm{p} , \ddot{L}_\mathrm{p} ,\ \mbox{higher
derivatives of}\ L_\mathrm{p} \right)$ and $L_\mathrm{IR} =
L_\mathrm{IR} \left(L_\mathrm{f} , \dot{L}_\mathrm{f} , \ddot{L}_\mathrm{f},\ 
\mbox{higher derivatives of}\ L_\mathrm{f} \right)$ as follows,
\begin{equation}
\frac{c^2}{\left(L_\mathrm{IR}\right)^2} 
= -\frac{\left(F \left(T^{\left(L_\mathrm{p} \right)} \right) 
 - T^{\left(L_\mathrm{p} \right)}\right)}{6}
 - 2\left(\frac{\dot{L}_p}{L_\mathrm{p}} - \frac{1}{L_\mathrm{p}}\right)^2~\left. 
 \frac{dF}{dT} \right|_{T=T^{\left(L_\mathrm{p} \right)}} \, ,
\label{hc_F(T)_ph}
\end{equation}
and
\begin{equation}
\frac{c^2}{\left(L_\mathrm{IR}\right)^2} = 
 -\frac{\left(F \left(T^{\left(L_\mathrm{f} \right)} \right) 
 - T^{\left(L_\mathrm{f} \right)}\right)}{6}
 - 2\left(\frac{\dot{L}_\mathrm{f}}{L_\mathrm{f}} + \frac{1}{L_\mathrm{f}}\right)^2~\left. 
 \frac{dF}{dT}\right|_{T=T^{\left(L_\mathrm{f} \right)}}
\, ,
\label{hc_F(T)_fh}
\end{equation}
respectively. The holographic cut-offs determined in
Eqs.~(\ref{hc_F(T)_ph}) and (\ref{hc_F(T)_fh}) constitute the
cosmological field equations and thus can provide an equivalent
holographic model for any arbitrary $F(T)$ gravity model. As an
example, we may consider \cite{Cai:2015emx},
\begin{equation}
F(T) = T - \alpha\left(-T\right)^p \, ,
\label{example2_F(T)}
\end{equation}
with $\alpha$ being a model parameter having mass dimension
$[2-2p]$ and $p$ is a dimensionless quantity. Earlier it was shown
that the $F(T)$ model in Eq.~(\ref{example2_F(T)}) (along with
suitable initial conditions) allow the Universe to evolve from an
initial phase of radiation domination to a cosmic acceleration at
late times for $p \neq 1$ \cite{Cai:2015emx}. With the help of
Eqs.~(\ref{hc_F(T)_ph}) and (\ref{hc_F(T)_fh}), we determine two
different forms of holographic cut-off for the model
(\ref{example2_F(T)}) as,
\begin{align}
\frac{c^2}{\left(L_\mathrm{IR}\right)^2}=&6^{p-1}\alpha(1 + 2p)
\left(\frac{\dot{L}_\mathrm{p}}{L_\mathrm{p}} - \frac{1}{L_\mathrm{p}}\right)^{2p}
 - 2\left(\frac{\dot{L}_\mathrm{p}}{L_\mathrm{p}} 
 - \frac{1}{L_\mathrm{p}}\right)^2 \nonumber\\
=&6^{p-1} \alpha (1 + 2p)\left(\frac{\dot{L}_\mathrm{f}}{L_\mathrm{f}} 
+ \frac{1}{L_\mathrm{f}}\right)^{2p}
 - 2\left(\frac{\dot{L}_\mathrm{f}}{L_\mathrm{f}} 
 + \frac{1}{L_\mathrm{f}}\right)^2 \, .
\label{example2_F(T)_hc}
\end{align}
The first line in the above equation gives the cut-off in terms of
the particle horizon and its derivatives, while the second line
gives the same however in terms of $L_\mathrm{f}$ and its
derivatives. Clearly the holographic energy density with the
$L_\mathrm{IR}$ of Eq.~(\ref{example2_F(T)_hc}) reproduces the
cosmological field equations for the model (\ref{example2_F(T)})
and hence drives the late-time accelerating epoch of our Universe.

\section{Conclusion}

In this paper, we applied the holographic principle to describe
the early and late-time acceleration epochs of our Universe in a
unified manner. Although holographic energy density has been well
studied at late times and recently it has also been applied in
inflation studies, giving rise to holographic dark energy and
inflationary realization respectively, however, to date it has not been
incorporated to unify various cosmological epochs of the Universe.
Such ``holographic unification'' is demonstrated in the present
paper, in the context of $F(R)$ and $f(\mathcal{G})$ gravity
theory without/with matter fields, where the corresponding
holographic cut-offs ($L_\mathrm{IR}$) are determined in terms of
the particle horizon and its derivatives or the future horizon and
its derivatives. For this purpose, we first prove the holographic
correspondence for general $F(R)$ or $f(\mathcal{G})$ theory and
then consider several specific forms of $F(R)$ or $f(\mathcal{G})$
(which are known to be viable models as per the unification of
inflation with the dark energy epoch is concerned) to show the
``holographic unification'' explicitly. One of the models
considered here is the axion-$F(R)$ gravity in presence of
radiation fluid, where the corresponding holographic energy
density that we propose is found to unify inflation with the
radiation, dark matter and dark energy epochs of the Universe in a
holographic context.

Moreover in the context of $F(R)$ gravity, apart from the two
aforementioned ways (where $L_\mathrm{IR}$ is determined in terms
of particle horizon or the future horizon), we also establish the
holographic cut-off in a different way, in particular, by an
integral form which along with $H = 1/L_\mathrm{IR}$ mimics the
cosmological dynamics of the corresponding model. The integral
form of $L_\mathrm{IR}$ has been discussed in the earlier
literature, however, these studies were focused on inflationary
models. Here we extended the determination of the integral form of
$L_\mathrm{IR}$ to the unified description of our Universe.

In summary, the holographic principle (where the cut-offs are in
terms of the particle horizon, or in terms of the future horizon
or in an integral form) proves to be very useful to unify the
cosmological eras of the Universe. However, our understanding for the choice of fundamental viable cut-off still 
remains to be lacking. The comparison of such cut-offs for realistic description of the universe evolution 
in unified manner may help in better understanding of holographic principle.

\begin{acknowledgments}
This work is partially supported  by the JSPS Grant-in-Aid for Scientific Research (C) 
No. 18K03615 (S.N.), and by MINECO (Spain), FIS2016-76363-P (S.D.O).
\end{acknowledgments}


\begin{thebibliography}{99}

%\cite{tHooft:1993dmi}
\bibitem{tHooft:1993dmi}
G.~'t Hooft,
%``Dimensional reduction in quantum gravity,''
Conf.\ Proc.\ C {\bf 930308} (1993) 284
[gr-qc/9310026].
%%CITATION = GR-QC/9310026;%%
%2258 citations counted in INSPIRE as of 15 Nov 2019

%\cite{Susskind:1994vu}
\bibitem{Susskind:1994vu}
L.~Susskind,
%``The World as a hologram,''
J.\ Math.\ Phys.\  {\bf 36} (1995) 6377
doi:10.1063/1.531249
[hep-th/9409089].
%%CITATION = doi:10.1063/1.531249;%%
%2761 citations counted in INSPIRE as of 15 Nov 2019

%\cite{Witten:1998qj}
\bibitem{Witten:1998qj}
E.~Witten,
%``Anti-de Sitter space and holography,''
Adv.\ Theor.\ Math.\ Phys.\  {\bf 2} (1998) 253
doi:10.4310/ATMP.1998.v2.n2.a2
[hep-th/9802150].
%%CITATION = doi:10.4310/ATMP.1998.v2.n2.a2;%%
%9761 citations counted in INSPIRE as of 15 Nov 2019

%\cite{Bousso:2002ju}
\bibitem{Bousso:2002ju}
R.~Bousso,
%``The Holographic principle,''
Rev.\ Mod.\ Phys.\  {\bf 74} (2002) 825
doi:10.1103/RevModPhys.74.825
[hep-th/0203101].
%%CITATION = doi:10.1103/RevModPhys.74.825;%%
%983 citations counted in INSPIRE as of 15 Nov 2019

%\cite{Cohen:1998zx}
%\bibitem{Cohen:1998zx}
%A.~G.~Cohen, D.~B.~Kaplan and A.~E.~Nelson,
%``Effective field theory, black holes, and the cosmological constant,''
%Phys.\ Rev.\ Lett.\  {\bf 82} (1999) 4971
%doi:10.1103/PhysRevLett.82.4971
%[hep-th/9803132].
%%CITATION = doi:10.1103/PhysRevLett.82.4971;%%
%918 citations counted in INSPIRE as of 15 Nov 2019

%\cite{Li:2004rb}
\bibitem{Li:2004rb}
M.~Li,
%``A Model of holographic dark energy,''
Phys.\ Lett.\ B {\bf 603} (2004) 1
doi:10.1016/j.physletb.2004.10.014
[hep-th/0403127].
%%CITATION = doi:10.1016/j.physletb.2004.10.014;%%
%1138 citations counted in INSPIRE as of 15 Nov 2019

%\cite{Wang:2016och}
\bibitem{Wang:2016och}
S.~Wang, Y.~Wang and M.~Li,
%``Holographic Dark Energy,''
Phys.\ Rept.\  {\bf 696} (2017) 1
doi:10.1016/j.physrep.2017.06.003
[arXiv:1612.00345 [astro-ph.CO]].
%%CITATION = doi:10.1016/j.physrep.2017.06.003;%%
%89 citations counted in INSPIRE as of 15 Nov 2019

%\cite{Pavon:2005yx}
\bibitem{Pavon:2005yx}
D.~Pavon and W.~Zimdahl,
%``Holographic dark energy and cosmic coincidence,''
Phys.\ Lett.\ B {\bf 628} (2005) 206
doi:10.1016/j.physletb.2005.08.134
[gr-qc/0505020].
%%CITATION = doi:10.1016/j.physletb.2005.08.134;%%
%436 citations counted in INSPIRE as of 15 Nov 2019

%\cite{Nojiri:2005pu}
\bibitem{Nojiri:2005pu}
S.~Nojiri and S.~D.~Odintsov,
%``Unifying phantom inflation with late-time acceleration: Scalar phantom-non-phantom transition model and generalized holographic dark energy,''
Gen.\ Rel.\ Grav.\  {\bf 38} (2006) 1285
doi:10.1007/s10714-006-0301-6
[hep-th/0506212].
%%CITATION = doi:10.1007/s10714-006-0301-6;%%
%555 citations counted in INSPIRE as of 15 Nov 2019


%\cite{Enqvist:2004xv}
\bibitem{Enqvist:2004xv}
K.~Enqvist and M.~S.~Sloth,
%``A CMB/dark energy cosmic duality,''
Phys.\ Rev.\ Lett.\  {\bf 93} (2004) 221302
doi:10.1103/PhysRevLett.93.221302
[hep-th/0406019].
%%CITATION = doi:10.1103/PhysRevLett.93.221302;%%
%124 citations counted in INSPIRE as of 15 Nov 2019

%\cite{Zhang:2005yz}
\bibitem{Zhang:2005yz}
X.~Zhang,
%``Statefinder diagnostic for holographic dark energy model,''
Int.\ J.\ Mod.\ Phys.\ D {\bf 14} (2005) 1597
doi:10.1142/S0218271805007243
[astro-ph/0504586].
%%CITATION = doi:10.1142/S0218271805007243;%%
%207 citations counted in INSPIRE as of 15 Nov 2019

%\cite{Guberina:2005fb}
\bibitem{Guberina:2005fb}
B.~Guberina, R.~Horvat and H.~Stefancic,
%``Hint for quintessence-like scalars from holographic dark energy,''
JCAP {\bf 0505} (2005) 001
doi:10.1088/1475-7516/2005/05/001
[astro-ph/0503495].
%%CITATION = doi:10.1088/1475-7516/2005/05/001;%%
%61 citations counted in INSPIRE as of 15 Nov 2019

%\cite{Elizalde:2005ju}
\bibitem{Elizalde:2005ju}
E.~Elizalde, S.~Nojiri, S.~D.~Odintsov and P.~Wang,
%``Dark energy: Vacuum fluctuations, the effective phantom phase, and holography,''
Phys.\ Rev.\ D {\bf 71} (2005) 103504
doi:10.1103/PhysRevD.71.103504
[hep-th/0502082].
%%CITATION = doi:10.1103/PhysRevD.71.103504;%%
%360 citations counted in INSPIRE as of 15 Nov 2019

%\cite{Ito:2004qi}
\bibitem{Ito:2004qi}
M.~Ito,
%``Holographic dark energy model with non-minimal coupling,''
Europhys.\ Lett.\  {\bf 71} (2005) 712
doi:10.1209/epl/i2005-10151-x
[hep-th/0405281].
%%CITATION = doi:10.1209/epl/i2005-10151-x;%%
%83 citations counted in INSPIRE as of 15 Nov 2019

%\cite{Gong:2004cb}
\bibitem{Gong:2004cb}
Y.~g.~Gong, B.~Wang and Y.~Z.~Zhang,
%``The Holographic dark energy revisited,''
Phys.\ Rev.\ D {\bf 72} (2005) 043510
doi:10.1103/PhysRevD.72.043510
[hep-th/0412218].
%%CITATION = doi:10.1103/PhysRevD.72.043510;%%
%94 citations counted in INSPIRE as of 15 Nov 2019

%\cite{Saridakis:2007cy}
\bibitem{Saridakis:2007cy}
E.~N.~Saridakis,
%``Restoring holographic dark energy in brane cosmology,''
Phys.\ Lett.\ B {\bf 660} (2008) 138
doi:10.1016/j.physletb.2008.01.004
[arXiv:0712.2228 [hep-th]].
%%CITATION = doi:10.1016/j.physletb.2008.01.004;%%
%117 citations counted in INSPIRE as of 15 Nov 2019

%\cite{Gong:2009dc}
\bibitem{Gong:2009dc}
Y.~Gong and T.~Li,
%``A Modified Holographic Dark Energy Model with Infrared Infinite Extra Dimension(s),''
Phys.\ Lett.\ B {\bf 683} (2010) 241
doi:10.1016/j.physletb.2009.12.040
[arXiv:0907.0860 [hep-th]].
%%CITATION = doi:10.1016/j.physletb.2009.12.040;%%
%46 citations counted in INSPIRE as of 15 Nov 2019

%\cite{BouhmadiLopez:2011xi}
\bibitem{BouhmadiLopez:2011xi}
M.~Bouhmadi-Lopez, A.~Errahmani and T.~Ouali,
%``The cosmology of an holographic induced gravity model with curvature effects,''
Phys.\ Rev.\ D {\bf 84} (2011) 083508
doi:10.1103/PhysRevD.84.083508
[arXiv:1104.1181 [astro-ph.CO]].
%%CITATION = doi:10.1103/PhysRevD.84.083508;%%
%20 citations counted in INSPIRE as of 15 Nov 2019

%\cite{Malekjani:2012bw}
\bibitem{Malekjani:2012bw}
M.~Malekjani,
%``Generalized holographic dark energy model described at the Hubble length,''
Astrophys.\ Space Sci.\  {\bf 347} (2013) 405
doi:10.1007/s10509-013-1522-2
[arXiv:1209.5512 [gr-qc]].
%%CITATION = doi:10.1007/s10509-013-1522-2;%%
%8 citations counted in INSPIRE as of 15 Nov 2019

%\cite{Khurshudyan:2014axa}
\bibitem{Khurshudyan:2014axa}
M.~Khurshudyan, J.~Sadeghi, R.~Myrzakulov, A.~Pasqua and H.~Farahani,
%``Interacting quintessence dark energy models in Lyra manifold,''
Adv.\ High Energy Phys.\  {\bf 2014} (2014) 878092
doi:10.1155/2014/878092
[arXiv:1404.2141 [gr-qc]].
%%CITATION = doi:10.1155/2014/878092;%%
%10 citations counted in INSPIRE as of 15 Nov 2019

%\cite{Landim:2015hqa}
\bibitem{Landim:2015hqa}
R.~C.~G.~Landim,
%``Holographic dark energy from minimal supergravity,''
Int.\ J.\ Mod.\ Phys.\ D {\bf 25} (2016) no.04,  1650050
doi:10.1142/S0218271816500504
[arXiv:1508.07248 [hep-th]].
%%CITATION = doi:10.1142/S0218271816500504;%%
%22 citations counted in INSPIRE as of 16 Nov 2019

%\cite{Gao:2007ep}
\bibitem{Gao:2007ep}
C.~Gao, F.~Wu, X.~Chen and Y.~G.~Shen,
%``A Holographic Dark Energy Model from Ricci Scalar Curvature,''
Phys.\ Rev.\ D {\bf 79} (2009) 043511
doi:10.1103/PhysRevD.79.043511
[arXiv:0712.1394 [astro-ph]].
%%CITATION = doi:10.1103/PhysRevD.79.043511;%%
%334 citations counted in INSPIRE as of 16 Nov 2019

%\cite{Li:2008zq}
\bibitem{Li:2008zq}
M.~Li, C.~Lin and Y.~Wang,
%``Some Issues Concerning Holographic Dark Energy,''
JCAP {\bf 0805} (2008) 023
doi:10.1088/1475-7516/2008/05/023
[arXiv:0801.1407 [astro-ph]].
%%CITATION = doi:10.1088/1475-7516/2008/05/023;%%
%89 citations counted in INSPIRE as of 16 Nov 2019

%\cite{Anagnostopoulos:2020ctz}
\bibitem{Anagnostopoulos:2020ctz}
F.~K.~Anagnostopoulos, S.~Basilakos and E.~N.~Saridakis,
%``Observational constraints on Barrow holographic dark energy,''
[arXiv:2005.10302 [gr-qc]].
%1 citations counted in INSPIRE as of 12 Jun 2020


%\cite{Zhang:2005hs}
\bibitem{Zhang:2005hs}
X.~Zhang and F.~Q.~Wu,
%``Constraints on holographic dark energy from Type Ia supernova observations,''
Phys.\ Rev.\ D {\bf 72} (2005) 043524
doi:10.1103/PhysRevD.72.043524
[astro-ph/0506310].
%%CITATION = doi:10.1103/PhysRevD.72.043524;%%
%348 citations counted in INSPIRE as of 16 Nov 2019

%\cite{Li:2009bn}
\bibitem{Li:2009bn}
M.~Li, X.~D.~Li, S.~Wang and X.~Zhang,
%``Holographic dark energy models: A comparison from the latest observational data,''
JCAP {\bf 0906} (2009) 036
doi:10.1088/1475-7516/2009/06/036
[arXiv:0904.0928 [astro-ph.CO]].
%%CITATION = doi:10.1088/1475-7516/2009/06/036;%%
%170 citations counted in INSPIRE as of 16 Nov 2019

%\cite{Feng:2007wn}
\bibitem{Feng:2007wn}
C.~Feng, B.~Wang, Y.~Gong and R.~K.~Su,
%``Testing the viability of the interacting holographic dark energy model by using combined observational constraints,''
JCAP {\bf 0709} (2007) 005
doi:10.1088/1475-7516/2007/09/005
[arXiv:0706.4033 [astro-ph]].
%%CITATION = doi:10.1088/1475-7516/2007/09/005;%%
%122 citations counted in INSPIRE as of 16 Nov 2019

%\cite{Zhang:2009un}
\bibitem{Zhang:2009un}
X.~Zhang,
%``Holographic Ricci dark energy: Current observational constraints, quintom feature, and the reconstruction of scalar-field dark energy,''
Phys.\ Rev.\ D {\bf 79} (2009) 103509
doi:10.1103/PhysRevD.79.103509
[arXiv:0901.2262 [astro-ph.CO]].
%%CITATION = doi:10.1103/PhysRevD.79.103509;%%
%118 citations counted in INSPIRE as of 16 Nov 2019

%\cite{Lu:2009iv}
\bibitem{Lu:2009iv}
J.~Lu, E.~N.~Saridakis, M.~R.~Setare and L.~Xu,
%``Observational constraints on holographic dark energy with varying gravitational constant,''
JCAP {\bf 1003} (2010) 031
doi:10.1088/1475-7516/2010/03/031
[arXiv:0912.0923 [astro-ph.CO]].
%%CITATION = doi:10.1088/1475-7516/2010/03/031;%%
%64 citations counted in INSPIRE as of 16 Nov 2019

%\cite{Micheletti:2009jy}
\bibitem{Micheletti:2009jy}
S.~M.~R.~Micheletti,
%``Observational constraints on holographic tachyonic dark energy in interaction with dark matter,''
JCAP {\bf 1005} (2010) 009
doi:10.1088/1475-7516/2010/05/009
[arXiv:0912.3992 [gr-qc]].
%%CITATION = doi:10.1088/1475-7516/2010/05/009;%%
%38 citations counted in INSPIRE as of 16 Nov 2019

%\cite{Huang:2004wt}
\bibitem{Huang:2004wt}
Q.~G.~Huang and Y.~G.~Gong,
%``Supernova constraints on a holographic dark energy model,''
JCAP {\bf 0408} (2004) 006
doi:10.1088/1475-7516/2004/08/006
[astro-ph/0403590].
%%CITATION = doi:10.1088/1475-7516/2004/08/006;%%
%278 citations counted in INSPIRE as of 16 Nov 2019

%\cite{Mukherjee:2017oom}
\bibitem{Mukherjee:2017oom}
P.~Mukherjee, A.~Mukherjee, H.~Jassal, A.~Dasgupta and N.~Banerjee,
%``Holographic dark energy: constraints on the interaction from diverse observational data sets,''
Eur. Phys. J. Plus \textbf{134} (2019) no.4, 147
doi:10.1140/epjp/i2019-12504-7
[arXiv:1710.02417 [astro-ph.CO]].
%3 citations counted in INSPIRE as of 01 May 2020

%\cite{Nojiri:2017opc}
\bibitem{Nojiri:2017opc}
S.~Nojiri and S.~Odintsov,
%``Covariant Generalized Holographic Dark Energy and Accelerating Universe,''
Eur. Phys. J. C \textbf{77} (2017) no.8, 528
doi:10.1140/epjc/s10052-017-5097-x
[arXiv:1703.06372 [hep-th]].
%41 citations counted in INSPIRE as of 01 May 2020

%\cite{Horvat:2011wr}
\bibitem{Horvat:2011wr}
R.~Horvat,
%``Holographic bounds and Higgs inflation,''
Phys. Lett. B \textbf{699} (2011), 174-176
doi:10.1016/j.physletb.2011.04.004
[arXiv:1101.0721 [hep-ph]].
%3 citations counted in INSPIRE as of 11 Jul 2020

%\cite{Nojiri:2019kkp}
\bibitem{Nojiri:2019kkp}
S.~Nojiri, S.~D.~Odintsov and E.~N.~Saridakis,
%``Holographic inflation,''
Phys. Lett. B \textbf{797} (2019), 134829
doi:10.1016/j.physletb.2019.134829
[arXiv:1904.01345 [gr-qc]].
%20 citations counted in INSPIRE as of 01 May 2020

%\cite{Paul:2019hys}
\bibitem{Paul:2019hys}
T.~Paul,
%``Holographic correspondence of $F(R)$ gravity with/without matter fields,''
EPL \textbf{127} (2019) no.2, 20004
doi:10.1209/0295-5075/127/20004
[arXiv:1905.13033 [gr-qc]].
%2 citations counted in INSPIRE as of 01 May 2020

%\cite{Bargach:2019pst}
\bibitem{Bargach:2019pst}
A.~Bargach, F.~Bargach, A.~Errahmani and T.~Ouali,
%``Induced gravity effect on inflationary parameters in an holographic cosmology,''
Int. J. Mod. Phys. D \textbf{29} (2020) no.02, 2050010
doi:10.1142/S0218271820500108
[arXiv:1904.06282 [hep-th]].
%0 citations counted in INSPIRE as of 01 May 2020

%\cite{Elizalde:2019jmh}
\bibitem{Elizalde:2019jmh}
E.~Elizalde and A.~Timoshkin,
%``Viscous fluid holographic inflation,''
Eur. Phys. J. C \textbf{79} (2019) no.9, 732
doi:10.1140/epjc/s10052-019-7244-z
[arXiv:1908.08712 [gr-qc]].
%0 citations counted in INSPIRE as of 01 May 2020

%\cite{Oliveros:2019rnq}
\bibitem{Oliveros:2019rnq}
A.~Oliveros and M.~A.~Acero,
%``Inflation driven by a holographic energy density,''
EPL \textbf{128} (2019) no.5, 59001
doi:10.1209/0295-5075/128/59001
[arXiv:1911.04482 [gr-qc]].
%0 citations counted in INSPIRE as of 01 May 2020

%\cite{Nojiri:2019yzg}
\bibitem{Nojiri:2019yzg}
S.~Nojiri, S.~D.~Odintsov and E.~N.~Saridakis,
%``Holographic bounce,''
Nucl. Phys. B \textbf{949} (2019), 114790
doi:10.1016/j.nuclphysb.2019.114790
[arXiv:1908.00389 [gr-qc]].
%5 citations counted in INSPIRE as of 01 May 2020

%\cite{Lehners:2008vx}
\bibitem{Lehners:2008vx}
J.~L.~Lehners,
%``Ekpyrotic and Cyclic Cosmology,''
Phys.\ Rept.\ {\bf 465} (2008) 223
doi:10.1016/j.physrep.2008.06.001
[arXiv:0806.1245 [astro-ph]].
%%CITATION = doi:10.1016/j.physrep.2008.06.001;%%
%146 citations counted in INSPIRE as of 23 Dec 2016

%\cite{Quintin:2014oea}
\bibitem{Quintin:2014oea}
J.~Quintin, Y.~F.~Cai and R.~H.~Brandenberger,
%``Matter creation in a nonsingular bouncing cosmology,''
Phys.\ Rev.\ D {\bf 90} (2014) no.6, 063507
doi:10.1103/PhysRevD.90.063507
[arXiv:1406.6049 [gr-qc]].
%%CITATION = doi:10.1103/PhysRevD.90.063507;%%
%44 citations counted in INSPIRE as of 23 Dec 2016

%\cite{Cai:2013vm}
\bibitem{Cai:2013vm}
Y.~F.~Cai, R.~Brandenberger and P.~Peter,
%``Anisotropy in a Nonsingular Bounce,''
Class.\ Quant.\ Grav.\ {\bf 30} (2013) 075019
doi:10.1088/0264-9381/30/7/075019
[arXiv:1301.4703 [gr-qc]].
%%CITATION = doi:10.1088/0264-9381/30/7/075019;%%
%49 citations counted in INSPIRE as of 23 Dec 2016

%\cite{Cai:2016thi}
\bibitem{Cai:2016thi}
Y.~Cai, Y.~Wan, H.~Li, T.~Qiu and Y.~Piao,
%``The Effective Field Theory of nonsingular cosmology,''
JHEP \textbf{01} (2017), 090
doi:10.1007/JHEP01(2017)090
[arXiv:1610.03400 [gr-qc]].
%79 citations counted in INSPIRE as of 02 May 2020

%\cite{Elizalde:2020zcb}
\bibitem{Elizalde:2020zcb}
E.~Elizalde, S.~Odintsov, V.~Oikonomou and T.~Paul,
%``Extended matter bounce scenario in ghost free $f(R,\mathcal{G})$ gravity compatible with GW170817,''
Nucl. Phys. B \textbf{954} (2020), 114984
doi:10.1016/j.nuclphysb.2020.114984
[arXiv:2003.04264 [gr-qc]].
%3 citations counted in INSPIRE as of 15 Jun 2020


%\cite{Brevik:2019mah}
\bibitem{Brevik:2019mah}
I.~Brevik and A.~Timoshkin,
%``Viscous Fluid Holographic Bounce,''
doi:10.1142/S0219887820500231
[arXiv:1911.09519 [gr-qc]].
%1 citations counted in INSPIRE as of 01 May 2020

%\cite{Coriano:2019eif}
\bibitem{Coriano:2019eif}
C.~Corianno and P.~H.~Frampton,
%``Holographic Principle, Cosmological Constant and Cyclic Cosmology,''
Mod. Phys. Lett. A \textbf{35} (2019) no.02, 1950355
doi:10.1142/S0217732319503553
[arXiv:1906.10090 [gr-qc]].
%0 citations counted in INSPIRE as of 01 May 2020

%\cite{Nojiri:2010wj}
\bibitem{Nojiri:2010wj}
S.~Nojiri and S.~D.~Odintsov,
%``Unified cosmic history in modified gravity: from $F(R)$ theory to Lorentz non-invariant models,''
Phys.\ Rept.\  {\bf 505} (2011) 59
doi:10.1016/j.physrep.2011.04.001
[arXiv:1011.0544 [gr-qc]].
%%CITATION = doi:10.1016/j.physrep.2011.04.001;%%
%2200 citations counted in INSPIRE as of 02 May 2020

%\cite{Nojiri:2017ncd}
\bibitem{Nojiri:2017ncd}
S.~Nojiri, S.~D.~Odintsov and V.~K.~Oikonomou,
%``Modified Gravity Theories on a Nutshell: Inflation, Bounce and Late-time Evolution,''
Phys.\ Rept.\  {\bf 692} (2017) 1
doi:10.1016/j.physrep.2017.06.001
[arXiv:1705.11098 [gr-qc]].
%%CITATION = doi:10.1016/j.physrep.2017.06.001;%%
%649 citations counted in INSPIRE as of 02 May 2020

%\cite{Capozziello:2011et}
\bibitem{Capozziello:2011et}
S.~Capozziello and M.~De Laurentis,
%``Extended Theories of Gravity,''
Phys.\ Rept.\  {\bf 509} (2011) 167
doi:10.1016/j.physrep.2011.09.003
[arXiv:1108.6266 [gr-qc]].
%%CITATION = doi:10.1016/j.physrep.2011.09.003;%%
%1364 citations counted in INSPIRE as of 16 Nov 2019

%\cite{Artymowski:2014gea}
\bibitem{Artymowski:2014gea}
M.~Artymowski and Z.~Lalak,
%``Inflation and dark energy from f(R) gravity,''
JCAP \textbf{09} (2014), 036
doi:10.1088/1475-7516/2014/09/036
[arXiv:1405.7818 [hep-th]].
%38 citations counted in INSPIRE as of 02 May 2020

%\cite{Nojiri:2003ft}
\bibitem{Nojiri:2003ft}
S.~Nojiri and S.~D.~Odintsov,
%``Modified gravity with negative and positive powers of the curvature: Unification of the inflation and of the cosmic acceleration,''
Phys. Rev. D \textbf{68} (2003), 123512
doi:10.1103/PhysRevD.68.123512
[arXiv:hep-th/0307288 [hep-th]].
%1529 citations counted in INSPIRE as of 02 May 2020

%\cite{Odintsov:2019mlf}
\bibitem{Odintsov:2019mlf}
S.~D.~Odintsov and V.~K.~Oikonomou,
%``$f(R)$ Gravity Inflation with String-Corrected Axion Dark Matter,''
Phys.\ Rev.\ D {\bf 99} (2019) no.6,  064049
doi:10.1103/PhysRevD.99.064049
[arXiv:1901.05363 [gr-qc]].
%%CITATION = doi:10.1103/PhysRevD.99.064049;%%
%26 citations counted in INSPIRE as of 02 May 2020

%\cite{Johnson:2019vwi}
\bibitem{Johnson:2019vwi}
J.~P.~Johnson and S.~Shankaranarayanan,
%``Low-energy modified gravity signatures on the large-scale structures,''
Phys. Rev. D \textbf{100} (2019) no.8, 083526
doi:10.1103/PhysRevD.100.083526
[arXiv:1904.07608 [astro-ph.CO]].
%0 citations counted in INSPIRE as of 02 May 2020

%\cite{Pinto:2018rfg}
\bibitem{Pinto:2018rfg}
P.~Pinto, L.~Del Vecchio, L.~Fatibene and M.~Ferraris,
%``Extended cosmology in Palatini f(R)-theories,''
JCAP \textbf{11} (2018), 044
doi:10.1088/1475-7516/2018/11/044
[arXiv:1807.00397 [gr-qc]].
%3 citations counted in INSPIRE as of 02 May 2020

%\cite{Odintsov:2019evb}
\bibitem{Odintsov:2019evb}
S.~D.~Odintsov and V.~K.~Oikonomou,
%``Unification of Inflation with Dark Energy in $f(R)$ Gravity and Axion Dark Matter,''
Phys.\ Rev.\ D {\bf 99} (2019) no.10,  104070
doi:10.1103/PhysRevD.99.104070
[arXiv:1905.03496 [gr-qc]].
%%CITATION = doi:10.1103/PhysRevD.99.104070;%%
%13 citations counted in INSPIRE as of 02 May 2020

%\cite{Nojiri:2019riz}
\bibitem{Nojiri:2019riz}
S.~Nojiri, S.~D.~Odintsov and V.~K.~Oikonomou,
%``$F(R)$ Gravity with an Axion-like Particle: Dynamics, Gravity Waves, Late and Early-time Phenomenology,''
arXiv:1907.01625 [gr-qc].
%%CITATION = ARXIV:1907.01625;%%
%6 citations counted in INSPIRE as of 02 May 2020

%\cite{Nojiri:2019fft}
\bibitem{Nojiri:2019fft}
S.~Nojiri, S.~D.~Odintsov and V.~K.~Oikonomou,
%``Unifying Inflation with Early and Late-time Dark Energy in $F(R)$ Gravity,''
arXiv:1912.13128 [gr-qc].
%%CITATION = ARXIV:1912.13128;%%
%3 citations counted in INSPIRE as of 02 May 2020

%\cite{Lobo:2008sg}
\bibitem{Lobo:2008sg}
F.~S.~Lobo,
%``The Dark side of gravity: Modified theories of gravity,''
[arXiv:0807.1640 [gr-qc]].
%222 citations counted in INSPIRE as of 02 May 2020

%\cite{Gorbunov:2010bn}
\bibitem{Gorbunov:2010bn}
D.~Gorbunov and A.~Panin,
%``Scalaron the mighty: producing dark matter and baryon asymmetry at reheating,''
Phys. Lett. B \textbf{700} (2011), 157-162
doi:10.1016/j.physletb.2011.04.067
[arXiv:1009.2448 [hep-ph]].
%68 citations counted in INSPIRE as of 02 May 2020

%\cite{Li:2007xn}
\bibitem{Li:2007xn}
B.~Li and J.~D.~Barrow,
%``The Cosmology of f(R) gravity in metric variational approach,''
Phys. Rev. D \textbf{75} (2007), 084010
doi:10.1103/PhysRevD.75.084010
[arXiv:gr-qc/0701111 [gr-qc]].
%263 citations counted in INSPIRE as of 02 May 2020

%\cite{Odintsov:2020nwm}
\bibitem{Odintsov:2020nwm}
S.~D.~Odintsov and V.~K.~Oikonomou,
%``Geometric Inflation and Dark Energy with Axion $F(R)$ Gravity,''
Phys.\ Rev.\ D {\bf 101} (2020) no.4,  044009
doi:10.1103/PhysRevD.101.044009
[arXiv:2001.06830 [gr-qc]].
%%CITATION = doi:10.1103/PhysRevD.101.044009;%%
%3 citations counted in INSPIRE as of 02 May 2020

%\cite{Odintsov:2020iui}
\bibitem{Odintsov:2020iui}
S.~D.~Odintsov and V.~K.~Oikonomou,
%``Aspects of Axion $F(R)$ Gravity,''
EPL {\bf 129} (2020) no.4,  40001
doi:10.1209/0295-5075/129/40001
[arXiv:2003.06671 [gr-qc]].
%%CITATION = doi:10.1209/0295-5075/129/40001;%%
%1 citations counted in INSPIRE as of 02 May 2020

%\cite{Appleby:2007vb}
\bibitem{Appleby:2007vb}
S.~A.~Appleby and R.~A.~Battye,
%``Do consistent $F(R)$ models mimic General Relativity plus $\Lambda$?,''
Phys. Lett. B \textbf{654} (2007), 7-12
doi:10.1016/j.physletb.2007.08.037
[arXiv:0705.3199 [astro-ph]].
%411 citations counted in INSPIRE as of 02 May 2020

%\cite{Elizalde:2010ts}
\bibitem{Elizalde:2010ts}
E.~Elizalde, S.~Nojiri, S.~Odintsov, L.~Sebastiani and S.~Zerbini,
%``Non-singular exponential gravity: a simple theory for early- and late-time accelerated expansion,''
Phys. Rev. D \textbf{83} (2011), 086006
doi:10.1103/PhysRevD.83.086006
[arXiv:1012.2280 [hep-th]].
%147 citations counted in INSPIRE as of 02 May 2020

%\cite{Cognola:2007zu}
\bibitem{Cognola:2007zu}
G.~Cognola, E.~Elizalde, S.~Nojiri, S.~Odintsov, L.~Sebastiani and S.~Zerbini,
%``A Class of viable modified f(R) gravities describing inflation and the onset of accelerated expansion,''
Phys. Rev. D \textbf{77} (2008), 046009
doi:10.1103/PhysRevD.77.046009
[arXiv:0712.4017 [hep-th]].
%510 citations counted in INSPIRE as of 02 May 2020

%\cite{Li:2007jm}
\bibitem{Li:2007jm}
B.~Li, J.~D.~Barrow and D.~F.~Mota,
%``The Cosmology of Modified Gauss-Bonnet Gravity,''
Phys. Rev. D \textbf{76} (2007), 044027
doi:10.1103/PhysRevD.76.044027
[arXiv:0705.3795 [gr-qc]].
%197 citations counted in INSPIRE as of 02 May 2020

%\cite{Odintsov:2018nch}
\bibitem{Odintsov:2018nch}
S.~D.~Odintsov, V.~K.~Oikonomou and S.~Banerjee,
%``Dynamics of inflation and dark energy from $F(R,G)$ gravity,''
Nucl.\ Phys.\ B {\bf 938} (2019) 935
doi:10.1016/j.nuclphysb.2018.07.013
[arXiv:1807.00335 [gr-qc]].
%%CITATION = doi:10.1016/j.nuclphysb.2018.07.013;%%
%16 citations counted in INSPIRE as of 02 May 2020

%\cite{Carter:2005fu}
\bibitem{Carter:2005fu}
B.~M.~Carter and I.~P.~Neupane,
%``Towards inflation and dark energy cosmologies from modified Gauss-Bonnet theory,''
JCAP \textbf{06} (2006), 004
doi:10.1088/1475-7516/2006/06/004
[arXiv:hep-th/0512262 [hep-th]].
%131 citations counted in INSPIRE as of 02 May 2020



%\cite{Nojiri:2019dwl}
\bibitem{Nojiri:2019dwl}
S.~Nojiri, S.~Odintsov, V.~Oikonomou, N.~Chatzarakis and T.~Paul,
%``Viable inflationary models in a ghost-free Gauss–Bonnet theory of gravity,''
Eur. Phys. J. C \textbf{79} (2019) no.7, 565
doi:10.1140/epjc/s10052-019-7080-1
[arXiv:1907.00403 [gr-qc]].
%6 citations counted in INSPIRE as of 12 Jun 2020


%\cite{Elizalde:2010jx}
\bibitem{Elizalde:2010jx}
E.~Elizalde, R.~Myrzakulov, V.~Obukhov and D.~Saez-Gomez,
%``LambdaCDM epoch reconstruction from F(R,G) and modified Gauss-Bonnet gravities,''
Class. Quant. Grav. \textbf{27} (2010), 095007
doi:10.1088/0264-9381/27/9/095007
[arXiv:1001.3636 [gr-qc]].
%132 citations counted in INSPIRE as of 02 May 2020

%\cite{Makarenko:2016jsy}
\bibitem{Makarenko:2016jsy}
A.~N.~Makarenko,
%``The role of Lagrange multiplier in Gauss–Bonnet dark energy,''
Int. J. Geom. Meth. Mod. Phys. \textbf{13} (2016) no.05, 1630006
doi:10.1142/S0219887816300063
%5 citations counted in INSPIRE as of 02 May 2020

%\cite{delaCruzDombriz:2011wn}
\bibitem{delaCruzDombriz:2011wn}
A.~de la Cruz-Dombriz and D.~Saez-Gomez,
%``On the stability of the cosmological solutions in $f(R,G)$ gravity,''
Class. Quant. Grav. \textbf{29} (2012), 245014
doi:10.1088/0264-9381/29/24/245014
[arXiv:1112.4481 [gr-qc]].
%61 citations counted in INSPIRE as of 02 May 2020

%\cite{Chakraborty:2018scm}
\bibitem{Chakraborty:2018scm}
S.~Chakraborty, T.~Paul and S.~SenGupta,
%``Inflation driven by Einstein-Gauss-Bonnet gravity,''
Phys. Rev. D \textbf{98} (2018) no.8, 083539
doi:10.1103/PhysRevD.98.083539
[arXiv:1804.03004 [gr-qc]].
%25 citations counted in INSPIRE as of 02 May 2020

%\cite{Kanti:2015pda}
\bibitem{Kanti:2015pda}
P.~Kanti, R.~Gannouji and N.~Dadhich,
%``Gauss-Bonnet Inflation,''
Phys. Rev. D \textbf{92} (2015) no.4, 041302
doi:10.1103/PhysRevD.92.041302
[arXiv:1503.01579 [hep-th]].
%60 citations counted in INSPIRE as of 02 May 2020

%\cite{Kanti:2015dra}
\bibitem{Kanti:2015dra}
P.~Kanti, R.~Gannouji and N.~Dadhich,
%``Early-time cosmological solutions in Einstein-scalar-Gauss-Bonnet theory,''
Phys. Rev. D \textbf{92} (2015) no.8, 083524
doi:10.1103/PhysRevD.92.083524
[arXiv:1506.04667 [hep-th]].
%31 citations counted in INSPIRE as of 02 May 2020

%\cite{Odintsov:2018zhw}
\bibitem{Odintsov:2018zhw}
S.~D.~Odintsov and V.~K.~Oikonomou,
%``Viable Inflation in Scalar-Gauss-Bonnet Gravity and Reconstruction from Observational Indices,''
Phys.\ Rev.\ D {\bf 98} (2018) no.4,  044039
doi:10.1103/PhysRevD.98.044039
[arXiv:1808.05045 [gr-qc]].
%%CITATION = doi:10.1103/PhysRevD.98.044039;%%
%24 citations counted in INSPIRE as of 02 May 2020

%\cite{Saridakis:2017rdo}
\bibitem{Saridakis:2017rdo}
E.~N.~Saridakis,
%``Ricci-Gauss-Bonnet holographic dark energy,''
Phys.\ Rev.\ D {\bf 97} (2018) no.6,  064035
doi:10.1103/PhysRevD.97.064035
[arXiv:1707.09331 [gr-qc]].
%%CITATION = doi:10.1103/PhysRevD.97.064035;%%
%13 citations counted in INSPIRE as of 16 Nov 2019

%\cite{Cognola:2006eg}
\bibitem{Cognola:2006eg}
G.~Cognola, E.~Elizalde, S.~Nojiri, S.~D.~Odintsov and S.~Zerbini,
%``Dark energy in modified Gauss-Bonnet gravity: Late-time acceleration and the hierarchy problem,''
Phys. Rev. D \textbf{73} (2006), 084007
doi:10.1103/PhysRevD.73.084007
[arXiv:hep-th/0601008 [hep-th]].
%506 citations counted in INSPIRE as of 02 May 2020



%\cite{Elizalde:2018rmz}
\bibitem{Elizalde:2018rmz}
E.~Elizalde, S.~D.~Odintsov, T.~Paul and D.~Chillon-Gomez,
%``Inflationary Universe in $F(R)$ gravity with antisymmetric tensor fields and their suppression during its evolution,''
Phys. Rev. D \textbf{99} (2019) no.6, 063506
doi:10.1103/PhysRevD.99.063506
[arXiv:1811.02960 [gr-qc]].
%16 citations counted in INSPIRE as of 02 May 2020

%\cite{Elizalde:2018now}
\bibitem{Elizalde:2018now}
E.~Elizalde, S.~Odintsov, V.~Oikonomou and T.~Paul,
%``Logarithmic-corrected $R^2$ Gravity Inflation in the Presence of Kalb-Ramond Fields,''
JCAP \textbf{02} (2019), 017
doi:10.1088/1475-7516/2019/02/017
[arXiv:1810.07711 [gr-qc]].
%15 citations counted in INSPIRE as of 02 May 2020


%\cite{Paul:2018jpq}
\bibitem{Paul:2018jpq}
T.~Paul and S.~SenGupta,
%``Dynamical suppression of spacetime torsion,''
Eur. Phys. J. C \textbf{79} (2019) no.7, 591
doi:10.1140/epjc/s10052-019-7109-5
[arXiv:1808.00172 [gr-qc]].
%5 citations counted in INSPIRE as of 12 Jun 2020



%\cite{Das:2018jey}
\bibitem{Das:2018jey}
A.~Das, T.~Paul and S.~Sengupta,
%``Invisibility of antisymmetric tensor fields in the light of $F(R)$ gravity,''
Phys. Rev. D \textbf{98} (2018) no.10, 104002
doi:10.1103/PhysRevD.98.104002
[arXiv:1804.06602 [hep-th]].
%5 citations counted in INSPIRE as of 12 Jul 2020

%\cite{Starobinsky:1980te}
\bibitem{Starobinsky:1980te}
A.~A.~Starobinsky,
%``A New Type of Isotropic Cosmological Models Without Singularity,''
Adv. Ser. Astrophys. Cosmol. \textbf{3} (1987), 130-133
doi:10.1016/0370-2693(80)90670-X
%4579 citations counted in INSPIRE as of 02 May 2020

%\cite{Capozziello:2002rd}
\bibitem{Capozziello:2002rd}
S.~Capozziello,
%``Curvature quintessence,''
Int.\ J.\ Mod.\ Phys.\ D {\bf 11} (2002) 483
doi:10.1142/S0218271802002025
[gr-qc/0201033].
%%CITATION = doi:10.1142/S0218271802002025;%%
%792 citations counted in INSPIRE as of 16 Nov 2019


%\cite{Odintsov:2018qug}
\bibitem{Odintsov:2018qug}
S.~D.~Odintsov, D.~Saez-Chillon Gomez and G.~S.~Sharov,
%``Testing logarithmic corrections on $R^2$-exponential gravity by observational data,''
Phys. Rev. D \textbf{99} (2019) no.2, 024003
doi:10.1103/PhysRevD.99.024003
[arXiv:1807.02163 [gr-qc]].
%6 citations counted in INSPIRE as of 01 Jun 2020

%\cite{Odintsov:2017qif}
\bibitem{Odintsov:2017qif}
S.~D.~Odintsov, D.~Sáez-Chillón Gómez and G.~S.~Sharov,
%``Is exponential gravity a viable description for the whole cosmological history?,''
Eur. Phys. J. C \textbf{77} (2017) no.12, 862
doi:10.1140/epjc/s10052-017-5419-z
[arXiv:1709.06800 [gr-qc]].
%27 citations counted in INSPIRE as of 01 Jun 2020



%\cite{Scolnic:2017caz}
\bibitem{Scolnic:2017caz}
D.~Scolnic, D.~Jones, A.~Rest, Y.~Pan, R.~Chornock, R.~Foley, M.~Huber, R.~Kessler, G.~Narayan, A.~Riess, S.~Rodney, E.~Berger, D.~Brout, P.~Challis, M.~Drout, D.~Finkbeiner, R.~Lunnan, R.~Kirshner, N.~Sanders, E.~Schlafly, S.~Smartt, C.~Stubbs, J.~Tonry, W.~Wood-Vasey, M.~Foley, J.~Hand, E.~Johnson, W.~Burgett, K.~Chambers, P.~Draper, K.~Hodapp, N.~Kaiser, R.~Kudritzki, E.~Magnier, N.~Metcalfe, F.~Bresolin, E.~Gall, R.~Kotak, M.~McCrum and K.~Smith,
%``The Complete Light-curve Sample of Spectroscopically Confirmed SNe Ia from Pan-STARRS1 and Cosmological Constraints from the Combined Pantheon Sample,''
Astrophys. J. \textbf{859} (2018) no.2, 101
doi:10.3847/1538-4357/aab9bb
[arXiv:1710.00845 [astro-ph.CO]].
%470 citations counted in INSPIRE as of 01 Jun 2020

%\cite{Wang:2016wjr}
\bibitem{Wang:2016wjr}
Y.~Wang \textit{et al.} [BOSS],
%``The clustering of galaxies in the completed SDSS-III Baryon Oscillation Spectroscopic Survey: tomographic BAO analysis of DR12 combined sample in configuration space,''
Mon. Not. Roy. Astron. Soc. \textbf{469} (2017) no.3, 3762-3774
doi:10.1093/mnras/stx1090
[arXiv:1607.03154 [astro-ph.CO]].
%53 citations counted in INSPIRE as of 01 Jun 2020

%\cite{Ade:2015xua}
\bibitem{Ade:2015xua}
P.~Ade \textit{et al.} [Planck],
%``Planck 2015 results. XIII. Cosmological parameters,''
Astron. Astrophys. \textbf{594} (2016), A13
doi:10.1051/0004-6361/201525830
[arXiv:1502.01589 [astro-ph.CO]].
%8693 citations counted in INSPIRE as of 01 Jun 2020

%\cite{Ratsimbazafy:2017vga}
\bibitem{Ratsimbazafy:2017vga}
A.~Ratsimbazafy, S.~Loubser, S.~Crawford, C.~Cress, B.~Bassett, R.~Nichol and P.~Väisänen,
%``Age-dating Luminous Red Galaxies observed with the Southern African Large Telescope,''
Mon. Not. Roy. Astron. Soc. \textbf{467} (2017) no.3, 3239-3254
doi:10.1093/mnras/stx301
[arXiv:1702.00418 [astro-ph.CO]].
%67 citations counted in INSPIRE as of 01 Jun 2020

%\cite{Marsh:2015xka}
\bibitem{Marsh:2015xka}
D.~J.~E.~Marsh,
%``Axion Cosmology,''
Phys. Rept. \textbf{643} (2016), 1-79
doi:10.1016/j.physrep.2016.06.005
[arXiv:1510.07633 [astro-ph.CO]].
%665 citations counted in INSPIRE as of 11 Jul 2020


%\cite{Ringwald:2016yge}
\bibitem{Ringwald:2016yge}
A.~Ringwald,
%``Alternative dark matter candidates: Axions,''
PoS \textbf{NOW2016} (2016), 081
doi:10.22323/1.283.0081
[arXiv:1612.08933 [hep-ph]].
%29 citations counted in INSPIRE as of 11 Jul 2020





%\cite{Akrami:2018odb}
\bibitem{Akrami:2018odb}
Y.~Akrami {\it et al.} [Planck Collaboration],
%``Planck 2018 results. X. Constraints on inflation,''
arXiv:1807.06211 [astro-ph.CO].
%%CITATION = ARXIV:1807.06211;%%
%558 citations counted in INSPIRE as of 16 Nov 2019

%\cite{Cai:2015emx}
\bibitem{Cai:2015emx}
Y.~Cai, S.~Capozziello, M.~De Laurentis and E.~N.~Saridakis,
%``f(T) teleparallel gravity and cosmology,''
Rept. Prog. Phys. \textbf{79} (2016) no.10, 106901
doi:10.1088/0034-4885/79/10/106901
[arXiv:1511.07586 [gr-qc]].
%417 citations counted in INSPIRE as of 02 May 2020

%\cite{Krssak:2015oua}
\bibitem{Krssak:2015oua}
M.~Kr\v{s}\v{s}\'{a}k and E.~N.~Saridakis,
%``The covariant formulation of f(T) gravity,''
Class. Quant. Grav. \textbf{33} (2016) no.11, 115009
doi:10.1088/0264-9381/33/11/115009
[arXiv:1510.08432 [gr-qc]].
%136 citations counted in INSPIRE as of 02 May 2020






\end{thebibliography}
\end{document}